\documentclass[a4paper,12pt]{article}
\usepackage[T1]{fontenc}
\usepackage{graphicx,caption}
\usepackage{amsmath}

\usepackage{amsfonts} 
\usepackage{amsthm} 
\usepackage{amssymb} 
\usepackage{array}
\usepackage[left=3cm,right=3cm,top=3cm,bottom=3cm]{geometry}

\usepackage[dvipsnames]{xcolor}
\usepackage{tikz}
\usetikzlibrary{shapes,arrows,positioning}
\usepackage[numbers]{natbib}

\usepackage{bm}

\usepackage{lipsum}
\usepackage{multirow}

\definecolor{vert}{rgb}{0.3,0.7,0.3}
\definecolor{bleu}{rgb}{0.4,0.6,1}
\definecolor{violet}{rgb}{0.2,0,0.7}
\definecolor{rouge}{rgb}{0.9,0.2,0.2}
\definecolor{violet2}{rgb}{0.7,0.4,0.7}

\def\tr{\ensuremath{\operatorname{tr}}}
\def\det{\ensuremath{\operatorname{det}}}
\newcommand{\ee}{{\rm e}}
\newcommand{\ii}{{\rm i}}
\newcommand{\dd}{{\rm d}}
\newcommand{\ep}{{\epsilon}}
\newcommand{\trans}{{\mathcal{T}}}
\newcommand{\cotan}{\mathrm{cotan}}
\newcommand{\bone}{\mathbf{b_1}}
\newcommand{\bL}{\mathbf{b_{N}}}
\newcommand{\vplus}{\mathbf{v_+}}
\newcommand{\vmoins}{\mathbf{v_-}}

\date{}

\begin{document}

\title{Topological edge states in two-gap unitary systems: A transfer matrix approach}

\author{Cl\'ement Tauber and Pierre Delplace\\ 
\footnotesize
  \it
   Laboratoire de Physique de l'\'Ecole Normale Sup\'erieure de Lyon,
   UMR CNRS 5672,\\ \footnotesize \it 46 all\'ee d'Italie, F-69364 LYON CEDEX 07, FRANCE}

\maketitle

\begin{abstract}
 We construct and investigate a family of two-band unitary systems living on a cylinder geometry and presenting
 localized edge states. Using the transfer matrix formalism, we solve and investigate in details such states
 in the thermodynamic limit. Analitycity considerations then suggest the construction of a family of 
 Riemman surfaces associated to the band structure of the system. In this picture, the corresponding edge 
 states naturally wind around non contractile loops, defining by the way a topological invariant associated
 to each gap of the system.
\end{abstract}

\section{Introduction}

First discovered in the context of the quantum Hall effect \cite{Laughlin1981, Halperin82, Buttiker88}, 
boundary states turn out to be the hallmark of topological properties that can emerge in any dimensions and 
for various symmetry classes \cite{KaneMele2005, schnyder08}.
It was also realized that these topological properties can be achieved in miscellaneous physical systems
beyond solids \cite{HasanKane10, QiZhang11} leading for instance to the discovery of chiral edge states in cold-atom \cite{Goldman13},
 electromagnetic \cite{HaldaneRaghu08, Hafezi11, Khanikaev13} and acoustic \cite{KaneLubensky14, Vitelli15, Witten15, Yang15} lattices.

These topological properties characterize the bulk bands and are encoded by a topological invariant
whose integer value cannot change unless bands touch, or equivalently, unless the gap closes.
For instance, in two dimensions and in absence of particular symmetry, this topological invariant is the first Chern number  \cite{Thouless82, HatsugaiPRL93}.
As the edge states live in the gaps spectrum, it follows that they cannot be removed or added unless a topological transition
of the bulk bands happens when the gap closes. This gives a topological robustness to the edge states.

Recently, similar topological properties have been found in \textit{unitary systems}, namely physical systems whose behavior is described 
by a unitary operator rather than a Hermitian operator. 
It follows that their spectrum is periodic, unlike an energy spectrum which is bounded.    
Among them are the Floquet systems that are periodic in time \cite{KitagawaPRB10, Lindner11}.
These dynamical systems, such like periodically driven solids \cite{Inoue10, Kitagawa11}, shaken cold-atom gases \cite{Jotzu14},
photonic lattices \cite{Fang12, Rechtsman2013} or discrete-time quantum walks 
\cite{Kitagawa12} are fully described by their (unitary) time evolution operator.
Importantly, it was shown that beyond the analogy with equilibrium (Hermitian) systems, 
they can exhibit edge states whereas the usual bulk topological invariants vanish.
This is understood as an exotic topological property characterized by a novel invariant assigned to a gap (rather than a band) and that accounts
for the time evolution over a period \cite{Rudner13, Carpentier15}.

However, there are unitary systems which are not Floquet systems but still exhibit edge states.
This is the case of photonic and microwave networks which can instead be described by (unitary) scattering matrices \cite{LiangPRL13, PasekChong14, HuPRX15}.
How to describe the topological origin of these edge states? Can it still be related to a bulk property?

To answer these questions we construct, in section \ref{sec:two-gap}, a generic model of a two-gap unitary system on a cylinder geometry.
 This model achieves the specific situation for which the topological invariants of the bands (namely the Chern number in our case) vanish. 
Then, in section \ref{sec:band structure}, we apply the transfer matrix method to investigate the appearance of edge states 
and discuss their simultaneous existence in the two gaps (thus guarantying a vanishing Chern number of each band).
In its seminal paper, Hatsugai showed that the transfer matrix approach provides a deep understanding of the topological nature of the edge states 
in the quantum Hall phase \cite{HatsugaiPRB93}. In particular, the transfer matrix allows one to focus directly on the gaps where the edge states live rather than the bands only.
We follow the same strategy and adapt this method to the unitary case in section \ref{sec:topo}. 
Our analysis reveals the underlying topologically non-trivial structure of the edge states and justifies the definition of a topological invariant assigned 
to a gap instead of a band. Finally, several examples of physical systems ruled by the present model are discussed in section \ref{sec:models}.

\section{Two-gap unitary models with topological edge states}
\label{sec:two-gap}

\subsection{Heuristic construction on a finite size lattice}

\begin{figure}[h]
\begin{tikzpicture}
\begin{scope}[yshift=2cm]
\draw[dashed] (-1,-1) -- (-1,-0.4);      \draw (-1,-0.4) -- (-1,2); \draw[dashed] (-1,2) -- (-1,2.6); 
\draw[dashed] (1,-1) -- (1,-0.4);        \draw (1,-0.4) -- (1,2);  \draw[dashed] (1,2) -- (1,2.6);
\draw[dashed] (-0.6,-1) -- (-0.6,-0.4);  \draw (-0.6,-0.4) -- (-0.6,2); \draw[dashed] (-0.6,2) -- (-0.6,2.6);
\draw[dashed] (-0.2,-1) -- (-0.2,-0.4);  \draw (-0.2,-0.4) -- (-0.2,2); \draw[dashed] (-0.2,2) -- (-0.2,2.6);
\draw[dashed] (0.2,-1) -- (0.2,-0.4);    \draw (0.2,-0.4) -- (0.2,2); \draw[dashed] (0.2,2) -- (0.2,2.6);
\draw[dashed] (0.6,-1) -- (0.6,-0.4);    \draw (0.6,-0.4) -- (0.6,2); \draw[dashed] (0.6,2) -- (0.6,2.6);

\draw (-1,2) -- (1,2);
\draw (-1,1.6) -- (1,1.6); 
\draw (-1,1.2) -- (1,1.2);
\draw (-1,0.8) -- (1,0.8);
\draw (-1,0.4) -- (1,0.4);
\draw (-1,0) -- (1,0);
\draw (-1,-0.4) -- (1,-0.4);
\draw[-latex] (-1.2,-2) -- (1.3,-2);
\draw (-1,-1.9)--(-1,-2.1)node[below]{$n=1$};
\draw (1,-1.9)--(1,-2.1)node[below]{$N$};

 \draw[-latex,violet,thick] (-1.1,2)--(-1.1,-0.5) ;
\draw[-latex,red,thick] (1.1,-0.5)--(1.1,2);
\end{scope}
\end{tikzpicture}
\qquad
\includegraphics[scale=0.4]{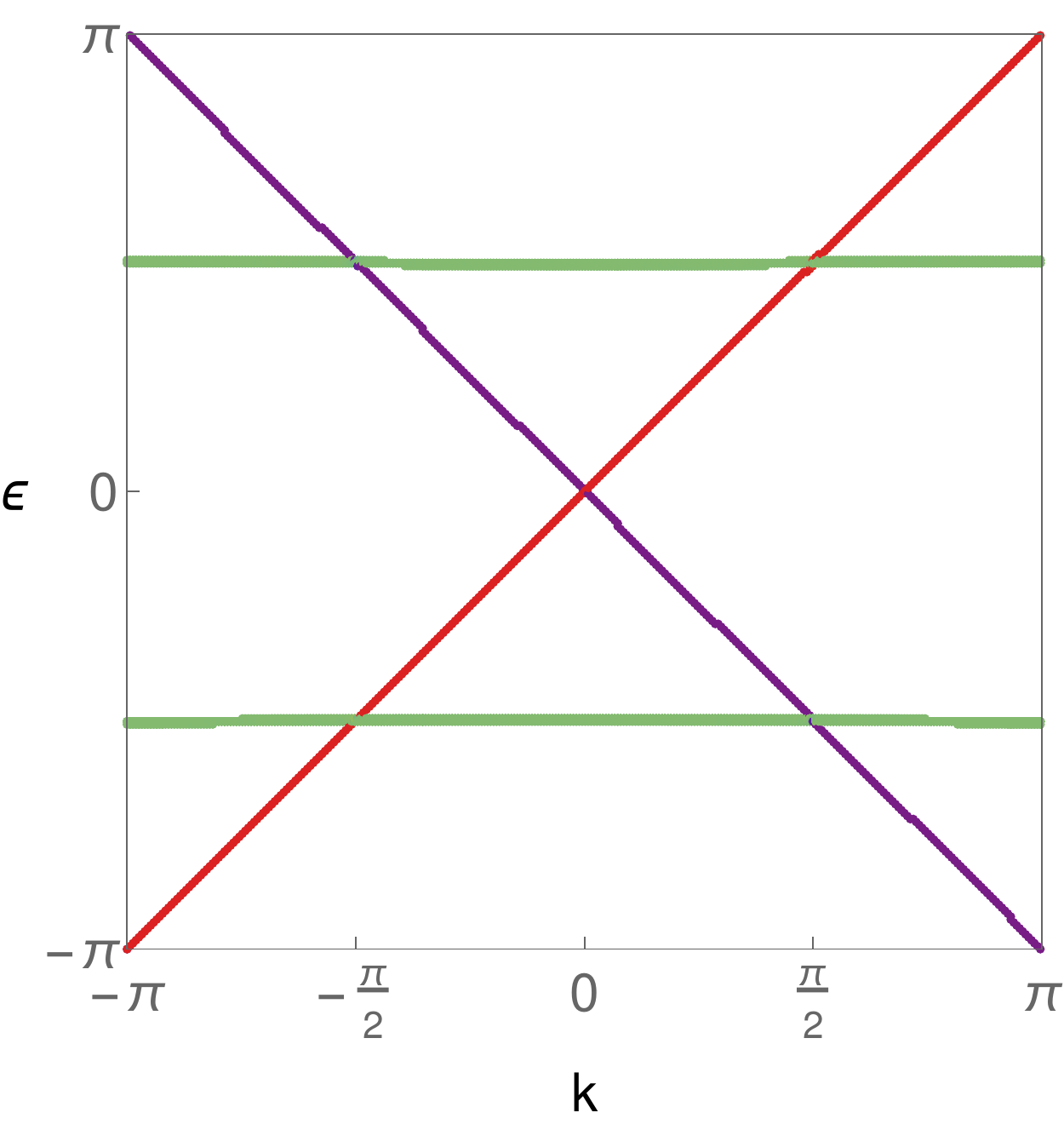}
\qquad
\includegraphics[scale=0.4]{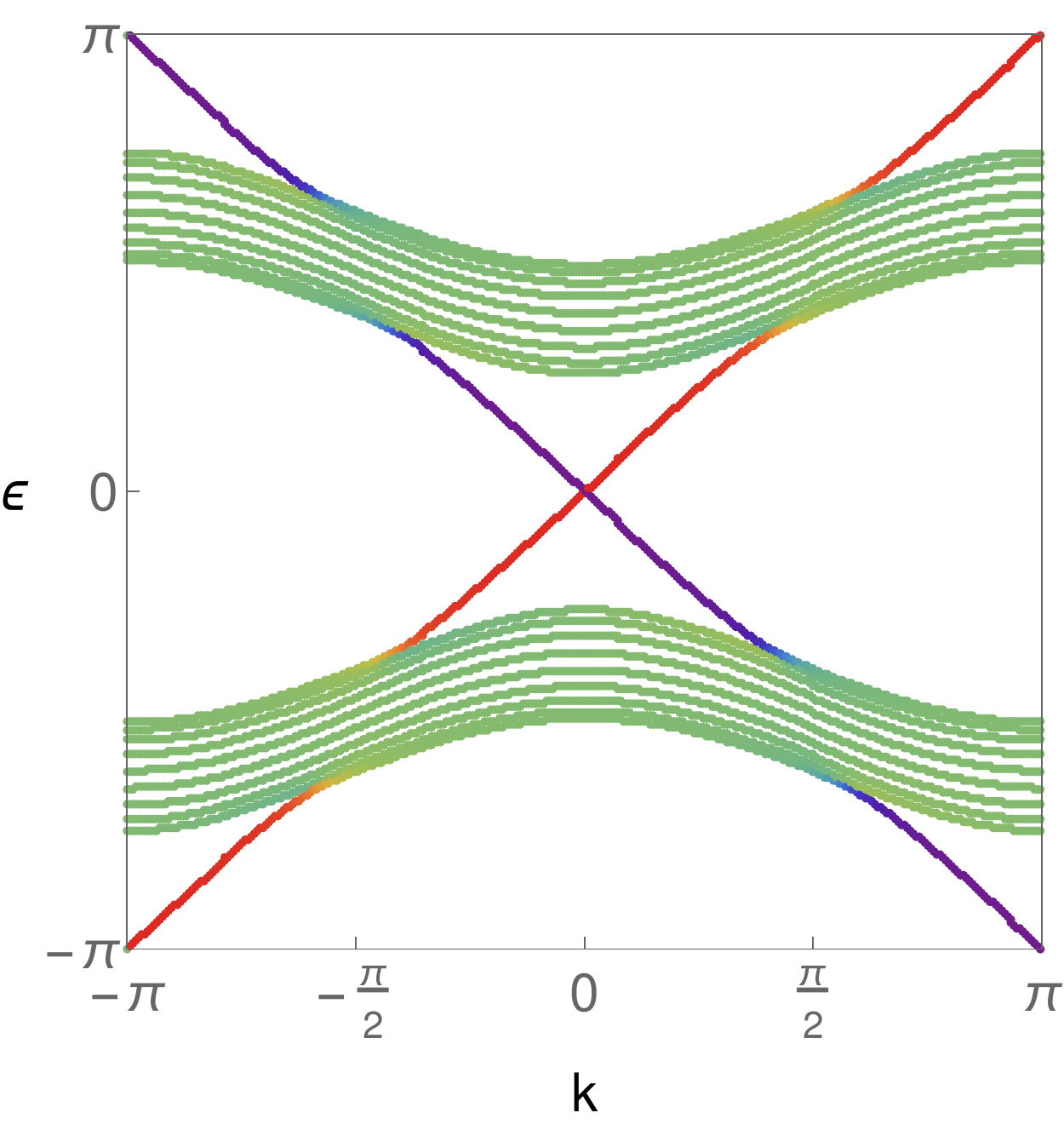}
\caption{\small (a) Sketch of the cylindrical geometry for the strip of width $N$.
The two arrows represent the chiral edge states at the boundary $n=1$ (purple) and $N$ (red).
Phase spectrum $\ep(k)$ obtained by a direct diagonalization of $\tilde{U}(k)$ with $N=10$ for $\theta=0$ (b) and $\theta=\pi/8$ (c).}
\label{fig:1}
\end{figure}

We  construct heuristically a two-gap unitary model for a strip geometry that exhibits topologically protected edge states.
For simplicity, we shall treat the cases of $0$ or $1$ edge state, but the generalization to several edge states is straightforward.  
To this end, we consider a system with two degrees of freedom -- that we refer to as $A$ and $B$ -- in the cylinder geometry 
such that the (dimensionless) quasi-momentum $k\in U(1)$ is a well-defined continuous parameter in the periodic direction,
whereas the lattice remains finite in the other one, as sketched in figure \ref{fig:1} (a). 
A state $|\Psi (k)\rangle$ of the system is then given by a $2N$-component vector $(A_1(k),B_1(k),\dots,A_N(k),B_N(k))^T$, which, 
for the scope of this study, is an eigenvector of a unitary matrix $\tilde{U}(k) \in U(2N)$, that is 
\begin{equation}
 \tilde{U}(k) |\Psi(k)\rangle = \ee^{-\ii \ep(k)} |\Psi(k)\rangle \ .
 \label{eq:Initial}
\end{equation}
We would like the phases $\ep(k)\in S_1$ of the eigenvalues of $\tilde{U}(2N)$ to display two-gaps and two edge states lying in these gaps.
Formally, the simplest way to obtain such a phase spectrum is to impose a quasi-diagonal form for $\tilde{U}(k)$ such as 
\begin{equation}
 \tilde{U}_0(k) = 
\begin{pmatrix}
        \ee^{\ii k} & & & \\
& \ii \sigma_y \otimes I_{N-1}  & &\\
&  & & \ee^{-\ii k}  
\end{pmatrix}
\label{eq:Utilde}
\end{equation} 
where the bulk part $\ii \sigma_y$ yields two flat bands $\ep=\pm\pi/2$ (in the range $\ep \in ]-\pi,\pi]$),
whereas $\ee^{ik}$ and $\ee^{-ik}$ guaranty the existence of propagating modes at the boundaries $A_1$ and $B_N$ respectively,
(see figure \ref{fig:1} (b)).
The phase spectrum $\ep(k)$ of $\tilde{U}_0(k)$ can then be obtained as a particular case of a more general unitary matrix 
\begin{equation}
\tilde U(k) \equiv 
\begin{pmatrix}
 1 & &  \\
  & U_2 \otimes I_{N-1} & \\
   & & 1
\end{pmatrix}
\cdot
\Big(\, U_1 \otimes I_N \,\Big)
\quad, \quad
 U_\mu = 
 \begin{pmatrix}
 \tau'_\mu  &  \rho'_\mu \\
 \rho_\mu &  \tau_\mu 
  \end{pmatrix}
\label{eq:Ufactorized} 
\end{equation}
where $U_\mu \in U(2)$ and $1$ is a scalar. One gets explicitly
\begin{equation}
\tilde U(k)= 
\left(
\begin{array}{ccccccccc}
 \scriptstyle\tau'_1 & \scriptstyle\rho'_1 & & & & & & & \\
 \hspace{-0.2cm}\scriptstyle\lceil  &   \multicolumn{2}{r}{_{\mathbf \Lambda}} & \quad \scriptstyle\rceil& & & & & \\
 \hspace{-0.2cm}\scriptstyle\lfloor  &  &    & \quad \scriptstyle\rfloor &&  & & & \\
& & & & & \scriptstyle\ddots & & & \\
& & & & & \hspace{-0.2cm} \scriptstyle\lceil & \multicolumn{2}{c}{\hspace{-0.2cm} _{\mathbf \Lambda}}  & \hspace{0.2cm} \scriptstyle\rceil \\
& & & & &  \hspace{-0.2cm}\scriptstyle\lfloor  &  & & \hspace{0.2cm}\scriptstyle\rfloor \\
& & & & & & &\scriptstyle\rho_1 &  \scriptstyle\tau_1
\end{array}\right)
,\quad  \mathbf \Lambda = \begin{pmatrix}
\scriptstyle \rho_1\tau'_2 &    \scriptstyle \tau_1\tau'_2  & \scriptstyle\tau'_1\rho'_2  & \scriptstyle\rho'_1 \rho'_2 \\
\scriptstyle\rho_1\rho_2&  \scriptstyle   \tau_1\rho_2   & \scriptstyle\tau'_1 \tau_2 & \scriptstyle\rho'_1 \tau_2
     \end{pmatrix}
\label{eq:Utilde_exp}
\end{equation} 
where, for concreteness, we choose
\begin{equation}
\tau'_1 =\tau_1^* = \cos{\theta}\, \ee^{\ii k}    \quad,\quad 
  \rho'_1=-\rho_1=\tau'_2 = \tau_2 = \sin{\theta} \quad,\quad
 \rho'_2 = -\rho_2 = \cos{\theta} \ .
\label{eq:model}
\end{equation}
In Eq. \eqref{eq:Utilde_exp}, $\Lambda$ is a $2\times 4$ matrix coupling components $A$ and $B$ at sites $n$ and $n+1$ for $1 \leq n < N$ (hence describing the bulk properties),
whereas first and last lines are constraints for $A$ and $B$ at sites $1$ and $N$ (hence yielding the boundary conditions). 

This model allows us to study the fate of the edge states when varying $\theta$. First, note
that $\tilde{U}(k)$ corresponds to $\tilde{U}_0(k)$ for $\theta=0$, up to a unitary transformation that does not change the spectrum. 
Then, for $\theta\neq 0$, the bulk bands acquire a dispersion (figure \ref{fig:1} (c)) and eventually 
touch at $(k=0,\ep=0)$ and $(k=\pi,\ep=\pi)$ for $\theta=\pi/4$, thus implying a transition toward a gapped phase with 
no edge state ($\pi/4<\theta<\pi/2$). 
A similar matrix $\tilde{U}(k)$ was derived to investigate the propagation of electromagnetic modes in arrays of optical resonators \cite{LiangPRL13, PasekChong14},
and can also me adapted to describe discrete-time quantum walks for a spin-$1/2$ particle or either a time-dependent tight-binding model, as we discuss in section \ref{sec:models}.

In the following, we investigate the topological properties of this model with the transfer matrix formalism.

\subsection{Transfer matrix formalism}

The transfer matrix formalism is a standard method to tackle a large variety of problems.
The starting point consists in relating the wave function amplitudes on adjacent sites of a lattice
by a matrix $\trans$. In our case, this translates as
\begin{equation}
\left(
 \begin{array}{c}
  A_{n+1}\\B_{n+1}
 \end{array}
 \right) = 
\trans
\left(
 \begin{array}{c}
  A_{n}\\B_{n}
 \end{array}
 \right) \ .
\label{eq:tranfer_def}
\end{equation}
From the general form of the matrix $\tilde{U}(k)$ in Eq. \eqref{eq:Utilde_exp}, one can infer the relation
\begin{equation}
\left(
 \begin{array}{cc}
 \tau'_1 \rho'_2  &  \rho'_1 \rho'_2 \\
 \tau'_1 \tau_2-\ee^{-i\ep} &  \rho'_1 \tau_2 
  \end{array}
 \right) 
\left(
 \begin{array}{c}
  A_{n+1}\\B_{n+1}
 \end{array}
 \right) = 
\left(
 \begin{array}{cc}
-\rho_1\tau'_2 & -\tau_1\tau'_2 + \ee^{-\ii \ep} \\
 -\rho_1\rho_2 &  -\tau_1\rho_2 
  \end{array}
 \right) 
\left(
 \begin{array}{c}
  A_{n}\\B_{n}
 \end{array}
 \right) \ .
 \label{eq:relation}
\end{equation} 
Note that the relation \eqref{eq:relation}, valid for $1\leq n<N$ describes the bulk; it does not take into account
the boundary conditions ($n=1,N$) that will be given in section \ref{sec:boundary}. 
We can then deduce the expression of $\trans$ which, for the specific parameters \eqref{eq:model}, reads
 \begin{align}
 \trans_\theta(k,\ep) = 
\left(
 \begin{array}{cc}
  \tan \theta  \, \ee^{\ii \ep}   &  -\ee^{\ii(\ep-k)} + \tan \theta  \\
-\ee^{\ii(\ep+k)} + \tan \theta   &  \cotan \theta \, \ee^{\ii\ep} + \frac{1}{\sin \theta \cos \theta} \, \ee^{-\ii \ep} -2\cos{k}
 \end{array}
 \right) \ .
 \label{eq:trans}
\end{align}

By construction, the matrix $\trans^l$ describes how the amplitude of a state evolves from the edge $n=1$ to a site $n=l$ of the bulk.
This information is then encoded into the eigenvalues $\lambda_{\pm}(\ep)$ of the transfer matrix $\trans$.
In particular, either $|\lambda_{\pm}(k,\ep)|=1$, which corresponds to an oscillatory (or delocalized) bulk state at $(k,\ep)$;
or one of the two eigenvalues satisfies $|\lambda(k,\ep)|<1$ -- note that $ \lambda_+ = \lambda_-^{-1}$ since $\det \trans =1$ 
-- and no delocalized state can appear i.e. there is a band gap. 
The eigenvalues of $\trans$ being solutions of the characteristic polynomial $ \lambda^2 - \tr \trans\, \lambda  + \det \trans = 0$,
they read
\begin{align}
\label{eq:lambda_ev}
f^2 > 1:  \qquad \lambda_{\pm}(k,\ep) &= -f \pm \sqrt{f^2-1}\ ,   &  \quad |\lambda_{\pm}(k,\ep)|\neq 1 \\
f^2 \leqslant 1:  \qquad \lambda_{\pm}(k,\ep) &= -f \pm \ii\sqrt{1-f^2}\ ,  &  \quad |\lambda_{\pm}(k,\ep)|= 1
\label{eq:lambda_bulk}
\end{align}
where $ f\equiv f_\theta(k,\ep) = -\frac{1}{2}\tr \trans(k,\ep)$ that is
\begin{equation}
 f  =  \cos{k} - \frac{\cos{\ep}}{\sin{\theta}\cos{\theta}} \ .
 \label{eq:f}
\end{equation}
In other words, for the model \eqref{eq:model}, the eigenvalues of the transfer matrix are fully determined by its trace. 
Finally, the two eigenvectors $\vmoins$ and $\vplus$, associated respectively to the eigenvalues $\lambda_-$ and $\lambda_+$ of the matrix 
$\trans = \left\{ \mathcal{T}_{ij}\right\}$, can be written as
\begin{equation}
 \mathbf{v_{\pm}}= \begin{pmatrix}
    \mathcal{T}_{12} \\
      \lambda_{\pm} - \mathcal{T}_{11}
   \end{pmatrix}  \ .
\label{eq:eigenvectors}
\end{equation}

In the following, we show how the projected bulk bands, the edge states and their topological properties can be inferred from
these eigenvalues and eigenvectors.

\section{Band structure from the transfer matrix}
\label{sec:band structure}

The transfer matrix formalism allows one to solve explicitly the initial problem $\eqref{eq:Initial}$,
namely to reconstruct the band structure and the corresponding states. 
We focus on the thermodynamic limit $N\rightarrow \infty$ of the infinite cylinder:
in that case the bulk bands become a continuum, whereas the edge states persist. 

\subsection{Bands and gaps}
\label{sec:bands and gaps}

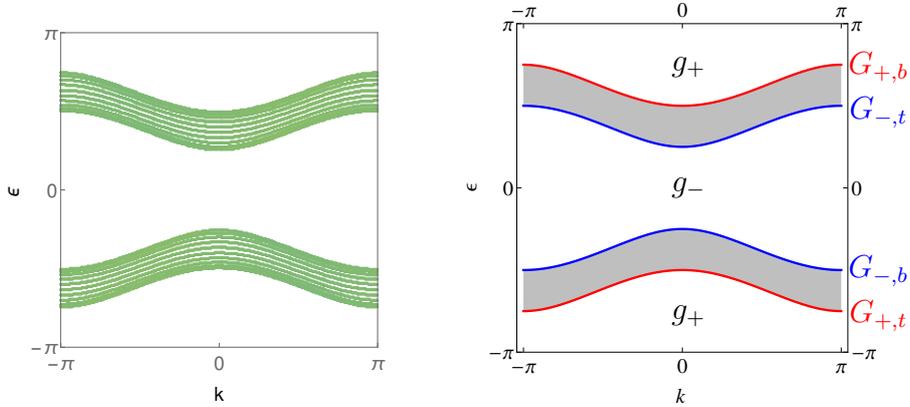
\begin{figure}[h]
\centering
\begin{tikzpicture}
\draw[blue] (5.8,1.2) node  {$G_{-,t}$};
\draw[red] (5.8,1.8) node  {$G_{+,b}$};
\draw[blue] (5.8,-0.9) node  {$G_{-,b}$};
\draw[red] (5.8,-1.5) node  {$G_{+,t}$};
\draw (3.3,0.2) node  {$g_-$};
\draw (3.3,1.8) node  {$g_+$};
\draw (3.3,-1.5) node  {$g_+$};
\node (0,0) {\includegraphics[scale=0.4]{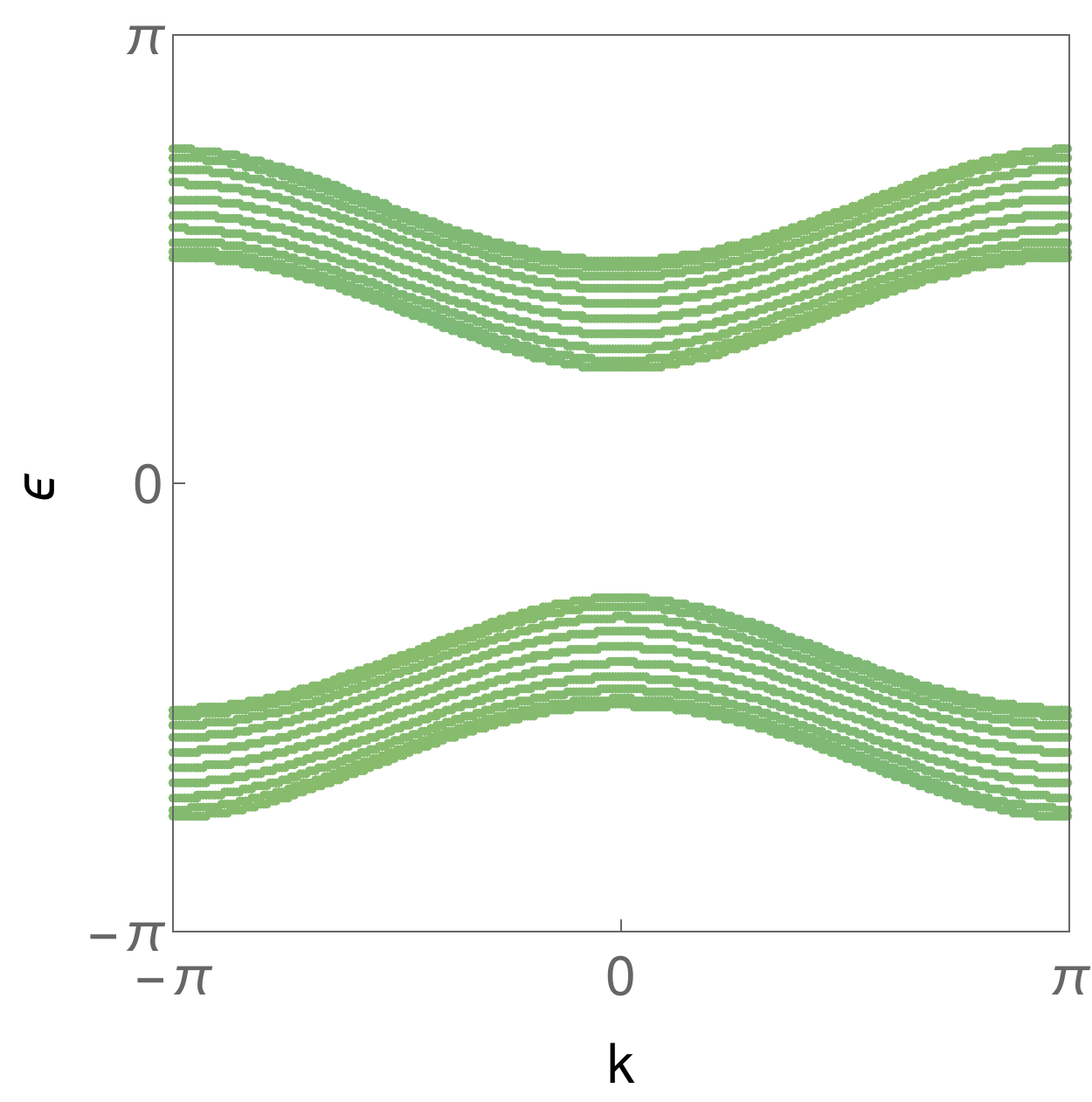}
\qquad\includegraphics[scale=0.6]{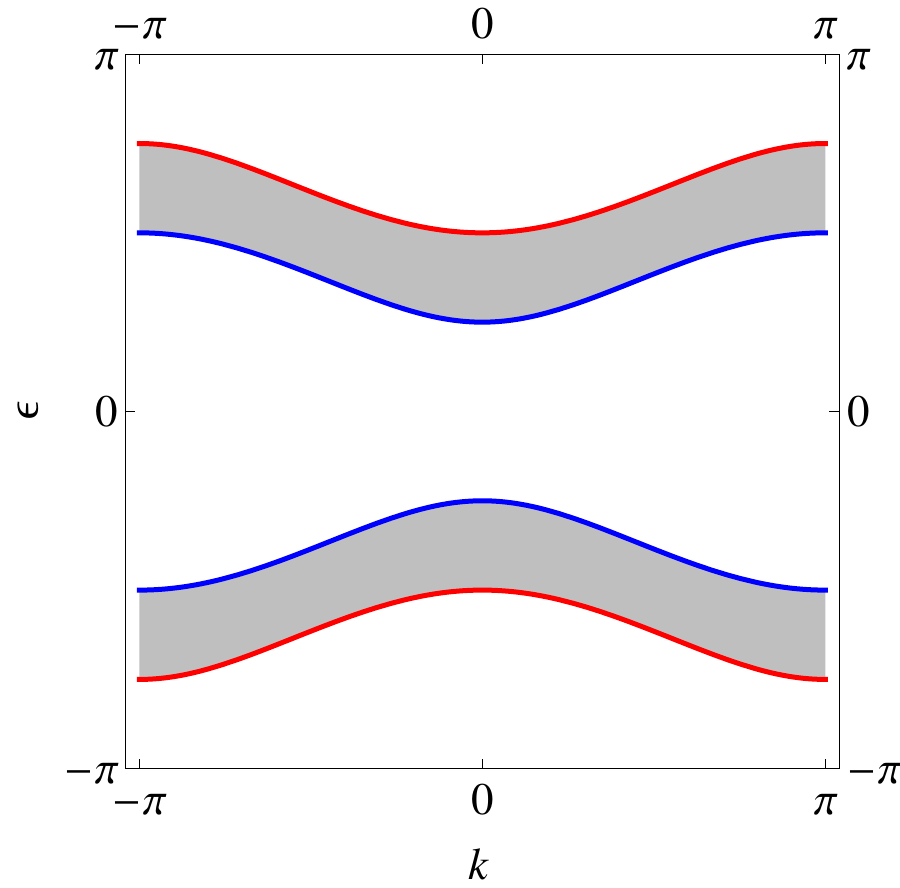}};
\end{tikzpicture}
\caption{\small Bulk bands of the strip obtained (left) from a direct diagonalization of $\tilde{U}(k)$ with $N=10$ and
(right) from the transfer matrix. The gap/band edges dispersions are represented in blue (red) for the gap $g_-$ ($g_+$).
The shaded area corresponds to the bulk band continuum at $N\rightarrow\infty$.}
\label{fig:bands}
\end{figure}

The bulk bands $\ep(k)$ correspond to regions of the diagram $(k,\ep$) for which
the eigenvalues of the transfer matrix satisfy $|\lambda_{\pm}(k,\ep)|=1$.
For the model we consider, this region is determined by the solutions of $f^2(k,\ep)\leqslant 1$ according to Eq. \eqref{eq:lambda_bulk}.
This can be found explicitly by using the expression of $f(k,\ep)$ in Eq. \eqref{eq:f}. 
The result is shown in figure \ref{fig:bands} and compared to a direct diagonalization of $\tilde{U}$.

Consistently, the gaps correspond to regions of the diagram $(k,\ep$) for which
the eigenvalues of the transfer matrix satisfy $|\lambda_{\pm}(k,\ep)|\neq1$, that is when $f^2(k,\ep)>1$. 
This inequality has two solutions $f(k,\ep)<-1$ and $f(k,\ep)>1$ yielding two different domains in the diagram $(k,\ep)$.
Since these two domains are separated by a band region ($f^2 \leqslant 1$), they thus correspond to the two gaped regions that we will
refer to as gap $g_-$ when $f<-1$ and $g_+$ when $f>1$.
The gaps $g_-$ and $g_+$ are respectively delimited by $f=-1$ or $f=+1$.
We therefore end up with a criteria that distinguishes the bands and the two gaps directly from the trace of the transfer matrix
\begin{equation}
 f <-1      \       \Leftrightarrow \ \text{gap $g_-$} \quad ,\quad  f^2 \leqslant 1 \  \Leftrightarrow  \ \text{bulk bands} 
   \quad , \quad f >1     \        \Leftrightarrow  \ \text{gap $g_+$}
  \end{equation}
The explicit dispersion relation of the two gap edges are then inferred from  Eq. (\ref{eq:f}) 
 \begin{equation}
  \begin{split}
  G_{-,t}(k)  = \arccos{\left(\sin\theta\cos\theta (\cos{k}+1)\right)} \quad , \quad G_{-,b}(k)  = -\arccos{\left(\sin\theta\cos\theta (\cos{k}+1)\right)}\\
  G_{+,b}(k)  = \arccos{\left(\sin\theta\cos\theta (\cos{k}-1)\right)} \quad , \quad G_{+,t}(k)  = -\arccos{\left(\sin\theta\cos\theta (\cos{k}-1)\right)}.
  \label{eq:gap_edge}
\end{split}
 \end{equation}
  and plotted in Fig. \ref{fig:bands}.

\subsection{Edge states}
\subsubsection{Boundary conditions}
\label{sec:boundary}

So far, the bulk part of $\tilde{U}$ have been encoded within the transfer matrix.
To characterize the existence of edge states, one also need to specify the boundary conditions.
Whereas in Hermitian systems, a standard boundary condition on a lattice consists in imposing the vanishing of the wavefunction,
such a procedure does not apply here since it is not compatible with unitarity.

As for the bulk transfer matrix, the boundary conditions are instead inferred from the original eigenvalues problem on $\tilde{U}$.
From Eq. \eqref{eq:Utilde_exp}, we relate for each edge $n=1,N$ the amplitudes $A_n$ and $B_n$ with the phase $\ep$ and the coefficients 
of $U_1$ as
\begin{equation}
\left(\tau'_1-\ee^{-\ii \ep}\right) A_1 = -\rho'_1 B_1 \qquad, \qquad  \rho_1 A_{N} = \left(\ee^{-\ii \ep}-\tau_1\right) B_{N} \ .
\end{equation}
A geometric way to reformulate these boundary conditions is to introduce two
 \textit{boundary vectors} $\bone$ and $\bL$ that are parallel to $\mathbf{a_1}=(A_1,B_1)$ and $\mathbf{a_{N}}=(A_{N},B_{N})$, that is
\begin{equation}
 \det (\mathbf{a_1},\bone) = 0 \qquad , \qquad \det (\mathbf{a_{N}},\bL) = 0 
\end{equation}
where we have defined 
\begin{equation}
 \bone = \left( \begin{array}{c}
                 -\rho'_1 \\
                  \tau'_1 - \ee^{-\ii \ep}
                \end{array}
                 \right)
\qquad , \qquad 
\bL   =   \left( \begin{array}{c}
                 \ee^{-\ii \ep}-\tau_1 \\
                  \rho_1 
                \end{array}
                 \right) \qquad .
\end{equation}
These two boundary vectors are related by the transfer matrix as $ \trans^{N-1} \bone \propto \bL $. 
This equation constrains the allowed values of couple $(k,\epsilon)$ for finite size $N$, thus yielding the corresponding 
solution of the initial problem\footnote{In particular this equation also gives the bulk bands dispersion relation at
finite size.} \eqref{eq:Initial}.
Next, we impose that in the thermodynamic limit ($N\rightarrow \infty$), $\bL$ does not depend on $N$ anymore.
Thus, in that limit, we can write $  \trans^{N} \bone =  \trans^{N-1} \bone$
so that $\bone$ becomes an eigenvector of the transfer matrix, given by \eqref{eq:eigenvectors}. 
Then, by decomposing $\bone$ as $\bone = \alpha \vplus + \beta \vmoins$, we end up with
\begin{equation}
 \alpha \lambda^{N-1}  \left( \lambda - 1\right)\, \vplus + \beta\lambda^{-(N-1)}\left(\lambda^{-1} - 1\right)\vmoins = 0 
\end{equation}
(where we have set $\lambda_+=\lambda$ for commodity).
 Let us assume in addition that $|\lambda|<1$.
 Then  $|\lambda|^{N-1}$ vanishes and $|\lambda^{-(N-1)}|$ tends to infinity.
It follows that $\beta=0$ and thus $\bone$ is proportional to $\vplus$. 
We can show in the same way that $\bL$ is proportional to $\vmoins$ when $N\rightarrow \infty$. 
This result will allow us to fully characterize the edge states in the next section.

Besides, $\mathbf{a_1}$ and $\mathbf{a_{N}}$ beeing also eigenstate of $\trans$ in that case, the corresponding state $ |\Psi \rangle$ of 
the initial problem is exponentially decreasing or increasing with typical length $\ln \lambda$, namely it is localized at one edge of the system.

\subsubsection{Existence and dispersion relation}
Concretely, in order to know which one of the $\lambda_{\pm}$'s satisfies $|\lambda|<1$, one has to fix $f$, which amounts to fix the gap.
Let us for instance focus on the gap $g_-$ (i.e. $f<-1$). 
Then, it is clear that $|\lambda_+|>1$ and thus $|\lambda_-|<1$.
 Therefore, in the gap $g_-$, an edge state at $n=1$ is given by the proportionality relation between the boundary vector $\bone$ 
and the eigenvector $\vmoins$, whereas the edge state at $n=N$ is given by the proportionality relation between
 the boundary vector $\bL$ and the eigenvector $\vplus$. 
The opposite situation occurs in the gap $g_+$, for which $f>1$, and thus $|\lambda_-|>1$ and $|\lambda_+|<1$, meaning that an edge state 
localized at the edge $n=1$ ($n=N$) is now associated with the eigenvalue $\lambda_+$ ($\lambda_-$) and the eigenvector $\vplus$ ($\vmoins$).
The existence of an each edge state and its dispersion relation is therefore given independently for each gap and for each boundary $n=1,N$
by the zeros of the function
\begin{equation}
  I_{n,\pm}(k,\ep) \equiv \det(\mathbf{b}_{n},\mathbf{v}_{\pm})
  \label{eq:det}
\end{equation}
as summarized in table \ref{t:table}.
\begin{table}[h]
\centering
\begin{tabular}{c|c|c}
& gap $g_-$ & gap $g_+$  \\
\hline
At edge\ $n=1$ & $I_{1,-}(k,\ep) = 0$ &   $I_{1,+}(k,\ep) = 0$   \\
 \hline
At edge\ $n=N$ & $I_{N,+}(k,\ep) = 0$    &   $I_{N,-}(k,\ep) = 0 $   \\
\end{tabular}
\caption{Equations for the existence of edge states in the space parameters ($k,\ep$).\label{t:table}}
\end{table}

\begin{figure}[htb]
\centering
\includegraphics[scale=0.23]{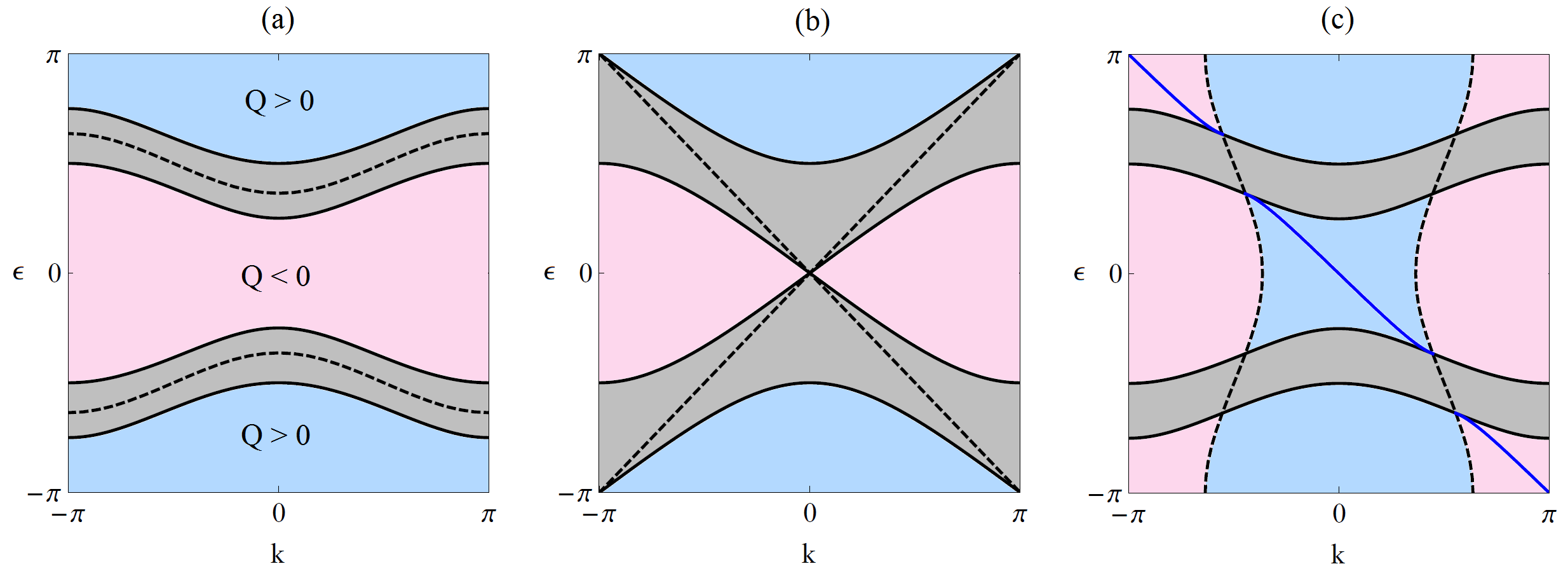}
\caption{\small Dispersion relations for (a) $\theta = 3\pi/8$, (b) $\theta = \pi/4$ and (c) $\theta = \pi/8$.
Blue area: $Q>0$ in the gap, pink area: $Q<0$ in the gap and dashed line: $Q=0$.
The edge state (represented here at boundary $n=1$) exists and touch a band at $k=\pm\arccos{(\pm \tan^2\theta)}$
when $Q$ changes sign.}
\label{fig:edgestates}
\end{figure}

Let us treat the case of the edge state $n=1$. Each of the two equations $I_{1,\pm}(k,\ep)=0$
 can be split into two equations corresponding to the real part and the imaginary part of 
$I_{1,\pm}(k,\ep)$. 
These two equations can be treated simultaneously, and after some algebra, one gets
\begin{eqnarray}
\label{eq:dispersion_edgestate}
&&\sin{\ep} = - \cos{\theta} \sin{k} \\
 Q(k,\ep) &\equiv &\frac{\cos{\ep}}{\cos{\theta}} - \frac{\cos{k}}{\sin{\theta}} = \pm \sqrt{f^2 - 1} \ .
 \label{eq:domain_edgestate}
 \end{eqnarray}
Whereas the explicit dispersion relation of the edge state localized at edge $n=1$ is straightforwardly obtained by inverting Eq. \eqref{eq:dispersion_edgestate}, 
its domain of validity in $k$ is constrained by the equation \eqref{eq:domain_edgestate} which depends on the gap $g_\pm$ through 
the contribution $\pm$ of the square root.
Concretely, the edge state only exists in the region $(k,\ep)$ satisfying by $Q(k,\ep)>0$ for the gap $g_+$ and in the region $Q(k,\ep)<0$ for the gap $g_-$.  
A similar procedure can be performed for the other edge so that the spectrum of the strip in the thermodynamic limit is recovered.
As displayed in figure \ref{fig:edgestates} this procedure provides the dispersion relation of the edge states (see also figure \ref{fig:shifted} (a) for the reconstruction of the full spectrum). 
Moreover, it shows that edge states cannot exist if the sign of $Q$ does not change in the gap. 
It reveals the importance of the role of the branch $\pm$ of the square root in Eq. \eqref{eq:domain_edgestate}. 
This is a crucial point to understand the topological robustness of the edge states
that will be developed in section \ref{sec:topo}. 

\subsubsection{Gaps correspondence}
\label{sec:gap_correspondence}

Before discussing the topological nature of the edge states, we emphasize that they always appear \textit{simultaneously} in the two gaps
for any generic unitary matrix $\tilde{U}$ of the form of Eq. \eqref{eq:Ufactorized} (provided the spectrum of $\tilde{U}$ is gapped). 
This is a particularly interesting property of unitary systems since the Chern number of the Bloch bands of the periodic system
in both $x$ and $y$ directions is guaranteed to vanish in this case \cite{Rudner13}. 
However, the conditions to achieve such a remarkable topological property are not clear and remain an open question.   
Here, we point out that the transfer matrix formalism is particularly useful to reveal 
a correspondence between the gaps that constrains the existence of an edge state simultaneously in each gap.

To make explicit such a property, we define a map $\upsilon$ acting in parameters space $\upsilon : (k,\ep) \rightarrow (\upsilon k, \upsilon \ep)$ that 
relates an edge state at the boundary $n=1$ that exists in the gap $g_+$ to an edge state localized at the same boundary but existing in the gap $g_-$.
We aim at determining under which conditions $I_{n,+}(k,\ep)$ and $I_{n,-}(\upsilon k, \upsilon \ep)$ vanish simultaneously.
By swapping the gaps, the function $f$ changes sign as $f(\upsilon k,\upsilon \ep) = - f(k,\ep)$ so that the eigenvalues become
\begin{equation}
  \lambda_{\pm}(k,\ep)\rightarrow \lambda_{\pm}(\upsilon k,\upsilon \ep) = - \lambda_{\mp} (k,\ep) \ .
  \label{eq:upsilon-lambda}
\end{equation}
It is clear that a map $\upsilon$ acting on the eigenvalues of the transfer matrix as in Eq. \eqref{eq:upsilon-lambda} actually changes the sign
of the trace of the transfer matrix. 
This allows us to express such a constraint on the transfer matrix itself as
\begin{equation}
   \sigma_z \, \trans(k,\ep)\, \sigma_z = - \Upsilon\, \trans(\upsilon k, \upsilon \ep) \Upsilon\
   \label{eq:Upsilon-trans}
\end{equation}
with $\sigma_z$ the standard Pauli matrix and where $\Upsilon$ is allowed to be either the complex conjugation operator $\kappa$ or the identity.
It is then easy to check that, for the model \eqref{eq:model}, the map $\upsilon : (k,\ep) \rightarrow (\pi- k, \pi- \ep)$
(defined modulo $2\pi$) satisfies Eq. \eqref{eq:Upsilon-trans} with $\Upsilon=\kappa$, so that Eq. \eqref{eq:upsilon-lambda} is satisfied as well.
It follows that an eigenvector $\mathbf{v_{\pm}}(k,\ep)$ of the transfer matrix $\trans(k,\ep)$ (see Eq. \eqref{eq:eigenvectors})
is transformed as 
\begin{equation}
 \mathbf{v_{\pm}} (\upsilon k,\upsilon \ep) = \sigma_z \kappa\, \mathbf{v}_{\mp}(k,\ep)
 \label{eq:constraint-eigenvector}
\end{equation}
 since $\mathcal{T}_{12}(\upsilon k, \upsilon\ep) = \mathcal{T}_{12}^*( k,\ep)$ 
and $\lambda_\pm(k,\ep) =\lambda_\pm^*(k,\ep)$ for the model \eqref{eq:model}. 

 A relation of proportionality between $I_{1,\pm}(k,\ep)$ and  
$I_{1,\mp}(\upsilon k,\upsilon \ep)$ can finally be inferred  provided that    
the boundary vector $\bone (k,\ep)$ -- being itself an eigenvector of the transfer matrix in the thermodynamic limit --  is also transformed as
\begin{equation}
 \bone (\upsilon k,\upsilon \ep) = \sigma_z \kappa\, \bone(k,\ep)
 \label{eq:constraint-bone}
\end{equation}
(as it can be checked explicitly from Eqs. \eqref{eq:trans} and \eqref{eq:eigenvectors}).
Indeed, according to Eqs. \eqref{eq:constraint-eigenvector} and \eqref{eq:constraint-bone} one gets    
\begin{align}
 I_{1,\pm} (k,\ep) &= \kappa \det \left(\sigma_z \bone, \sigma_z\mathbf{v}_{\mp} \right) (\upsilon k, \upsilon \ep) \\
 &= - \kappa\, \det \left( \bone, \mathbf{v}_{\mp} \right)(\upsilon k, \upsilon \ep) \\
 &= - \kappa\, I_{1,\mp}(\upsilon k, \upsilon \ep) \ .
\end{align}
and thus $I_{1,\pm} (k,\ep)$ and $I_{1,\mp}(\upsilon k, \upsilon \ep)$ vanish simultaneously.
This shows that if an edge state localized at edge $n=1$ exists in the gap $g_+$ (defined by the zeros of $I_{1,+}$), then so does an edge state
localized at the same edge in the gap $g_-$ (defined by the zeros of $I_{1,-}$).


\section{Topological property of the edge states from the transfer matrix}
\label{sec:topo}
\subsection{Riemann surface}

\subsubsection{Motivations} 

We would like to specify the topological nature of the edge states regardless the topological properties of the bands which are anyway always 
characterized by a vanishing Chern number as discussed above. 
A natural approach then consists in assigning a winding number to the edge states themselves \cite{Schuba13,GrafPorta13,AgazziEckmannGraf14}. 
When $\theta=0$, then $\tilde{U}(k)$ reduces to a block-diagonal matrix where the contributions of the bulk and of each edge can be distinguished.
In that case, we notice that each edge term $h_\pm(k)\equiv \ee^{\pm\ii k}$ is a $U(1)-$valued scalar so that its winding is well defined \cite{Rudner13}
and we get $\frac{1}{2\ii\pi}\int_{-\pi}^{\pi}dk\,h_\pm^\dagger \partial_k h_\pm =\pm1$.
How to construct such a quantity for arbitrary $\theta$ when $\tilde{U}(k)$ is not block-diagonal anymore? How to assign a specific gap to this winding? 
It is clear that the dispersion relations of the edge states obtained before will be a key ingredient.
However they are not defined for every $k$, not even periodic when considering one edge, and multivalued when considering both edges in one relation. 
Hence the winding number of such quantity is ill defined.

A possibility to overcome this problem is to consider instead the functions $I_{n,\pm}(k,\epsilon)$, whose zeros fully define 
the edge states dispersion relations. These functions depend explicitly on the eigenvalues $\lambda_\pm$ of the transfer matrix that contain the square root $\sqrt{f^2-1}$.
Therefore, their first derivative diverge around $f = \pm 1$, that is when $(k,\epsilon)$ touch the bands edges.
This divergence of the first derivative  appears in the wave function too, which -- remembering that the boundary vectors are 
also eigenvectors of the transfer matrix -- is proportional to $(A_n,B_n)^T \propto \lambda_\pm^{n-1} \mathbf{b}_{1,N}$.
Thus, to properly define a topological invariant, one first needs to find a way to obtain smooth solutions for the wave-functions when changing the branch
of the square root $\pm\sqrt{f^2-1}\rightarrow \mp \sqrt{f^2-1}$. 

The natural tool to deal with this issue is the theory of Riemann surfaces, 
as used by Hatsugai in 1993 to correctly define the winding number of the edge states in the (Hermitian) Harper model \cite{HatsugaiPRB93}.  
To construct the Riemann surfaces associated to our problem, we adapt the approach by Hatsugai to the unitary case where the eigenenergies  
are replaced by $\exp(-\ii \epsilon)$ which are $U(1)$-valued.
Substituting $f$ by its expression, this approach suggests to consider quantity 
$ \mu_k^2 = \left( \cos{k} - \frac{\cos{\ep}}{\sin{\theta}\cos{\theta}} \right)^2-1 $
for $\epsilon \mapsto z \in \mathbb C$ rather than the square root itself which is not analytic in the complex plan.
Such equation can be rewritten
\begin{equation}\label{eq:muQe}
 \mu_k^2 = \left( \dfrac{\ee^{\ii \epsilon}}{\sin 2\theta} \right)^2 \prod_{j=1}^4 \left( \ee^{-\ii \epsilon} - \phi_j(k) \right)
\end{equation}
where the expression of the four $\mathbb C$-valued functions $\phi_j(k)$ can be obtained explicitly.
Here we just need to know that $\phi_j(k) \in U(1)$, $\phi_j(k) \neq \phi_{j'}(k)$ for $j \neq j'$ 
(the gaps are not closing, except at $\theta =\pi/4$ which has to be avoided) 
and $\phi_j(k)$ is smooth in $k$ as soon as $0 < \theta < \pi/2$ 
(the bands are not degenerated to a line).
Note that when $\ee^{-\ii \epsilon} = \phi_j(k)$ one has $\mu_k = 0$ that is $f^2=1$: this corresponds to the gap/band edge.

The construction of the Riemann surface consists in two steps.
One is the standard construction of an elliptic curve and its geometric interpretation as a compact surface obtained by gluing two Riemann sheets
corresponding to the two branches $\pm$ of the square root. 
The second step, specific to unitary systems, accounts for the periodic structure of the phases
through the dependence in $\exp{\pm \ii \ep}$ (and its powers) appearing in $\mu_k^2$.

\subsubsection{Elliptic curves}

From now on, we consider a fixed momentum $k$ and forget about it in the notations.
We start with the standard construction of elliptic curves by considering the Riemann surface
\begin{equation}
 \mathcal R_0 = \left\lbrace (Z,\mu) \in \mathbb C^2 \, \left| \,  \mu^2 = W(Z)= \dfrac{1}{Z^2} \prod_{j=1}^4 (Z - \phi_j) \right. \right\rbrace
\end{equation}
with the $\phi_j$ satisfying the previous assumptions. 
This is an almost standard construction \cite{Bobenko11}.
First, this is a compact Riemann surface since it can be covered by open subset that are homeomorphic to open subset in $\mathbb C$,
with holomorphic transition function on the intersections. 
The singularities that appear when $Z \rightarrow 0$ and $\infty$ are not essential because 
the corresponding neighborhood can also be identified with open subsets of $\mathbb C$. 
There is a nice interpretation of this curve as a double covering of $\mathbb C$, or its compactified version \cite{Bobenko11}, as we
will see in the next section.

The complex square root cannot be defined analytically over the whole complex plane $\mathbb C$,
nor on its compact version, namely the Riemann sphere $\mathcal S  = \mathbb C \cup {\infty}$   \cite{Bobenko11}.
A natural branch cut on $\mathbb C$ (or $\mathcal S$) is defined such that $ W(Z) < 0$.
We know from Eq. \eqref{eq:muQe} that $W(Z)$ is real for $Z \in U(1)$ and that the sign of $W$ changes at $\phi_i$ where $\mu = 0$.
By construction, the $\phi_i$ are located on the equator of the Riemann sphere.
We chose the order of the $\phi_i$'s such that $W(Z) <0$ for $Z \in [\phi_1,\phi_2]$ and $Z \in [\phi_3,\phi_4]$ 
(this notation is ambiguous in $U(1)$ but that does not matter in the following). 
As a result $W(Z)$ is real and positive for $Z$ on the equator of each Riemann sphere as well, but outside the branch cuts.
To get an analytic structure for $\mathcal R_0$, consider two copies of such \textit{cut} Riemann spheres $\mathcal S^+$ and $\mathcal S^-$ 
and set the convention that $\mu = +\sqrt{W}>0$ (respectively $-\sqrt{W}<0$) on such a region of the Riemann sphere $\mathcal S^+$
($\mathcal S^-$).
Hence, by construction of these two copies,  one travels from one sphere to the other by crossing a branch cut, 
such that $\mu$ goes smoothly from positive to negative values  \cite{Bobenko11}.
The square root is now analytic over $\mathcal R_0$ (instead of $\mathcal S$) which is a torus of genus $1$ (instead of a sphere).
The Riemann surface  $\mathcal R_0$ is locally homeomorphic to the complex plane, but \textit{not globally} since the \textit{topology} is different.
This is the price to pay to have smooth functions. But this also gives a direct geometrical interpretation of the winding of the edge states 
\cite{HatsugaiPRB93}, as we will see.

\subsubsection{Punctured torus for the unitary problem}

\begin{figure}[htb]
 \centering
\begin{tikzpicture}
 \draw[blue,thick] (0,0) ellipse (1 and 0.5);
 \draw[dotted] (0,1.5) ellipse (1 and 0.5);
 \draw (-1,-2) -- (-1,2); \draw (1,-2) -- (1,2);
 \draw[-latex] (-1.2,-0.5) -- node[left]{$\eta$} (-1.2,0.5);
 \draw[-latex] (-120:1 and 0.7) arc (-120:-60:1 and 0.7);
 \draw[] (0,-0.7) node[below]{$\epsilon$};
 \draw[dashed] (-1,2) -- (-1,2.5); \draw[dashed] (1,2) -- (1,2.5); \draw[red] (0,2.5) node{$+\infty$};
  \draw[dashed] (-1,-2) -- (-1,-2.5); \draw[dashed] (1,-2) -- (1,-2.5); \draw[red] (0,-2.5) node{$-\infty$};
 \draw (1.5,-2) node {$\mathcal C$};

\draw[->,thick] (2.5,0) -- node[above]{$\varphi$} (4,0);

\begin{scope}[xshift=7cm]
 \draw[-latex] (-2.3,0) -- (2.3,0); \draw[-latex] (0,-2.3) -- (0,2.3);
 \draw[blue,thick] (0,0) circle(0.8);
 \draw[dotted] (0,0) circle(1.5);
 \draw[red,thick] (-0.1,-0.1) -- (0.1,0.1);\draw[red,thick] (0.1,-0.1) -- (-0.1,0.1); \draw[red] (0,0) node [below right]{$0$};
 \draw[red] (1.8,1.8) node {$\infty$};
 \draw (1.5,-2) node {$\mathbb C \setminus \{0,\infty\}$};
\end{scope}

\draw (10,0) node {$\cong$};

\begin{scope}[xshift=12cm]
\draw (0,0) circle (1.5);
\draw[blue,thick] (0,0) ellipse (1.5 and 0.5);
\draw[dotted] (0,1) ellipse (1.1 and 0.25);
\draw[red,thick] (-0.1,1.4) -- (0.1,1.6); \draw[red,thick] (0.1,1.4) -- (-0.1,1.6); \draw[red] (0,1.6) node[above]{$\infty$};
\draw[red,thick] (-0.1,-1.4) -- (0.1,-1.6); \draw[red,thick] (0.1,-1.4) -- (-0.1,-1.6); \draw[red] (0,-1.6) node[below]{$0$};
 \draw (1,-2) node {$\mathcal S^*$};
\end{scope}

\end{tikzpicture}
\caption{\small Sending the complex cylinder to the complex plane, 
or equivalently to the Riemann sphere with two forbidden points (essential singularities): $0$ and $\infty$. \label{fig:RiemSphere}}
\end{figure}
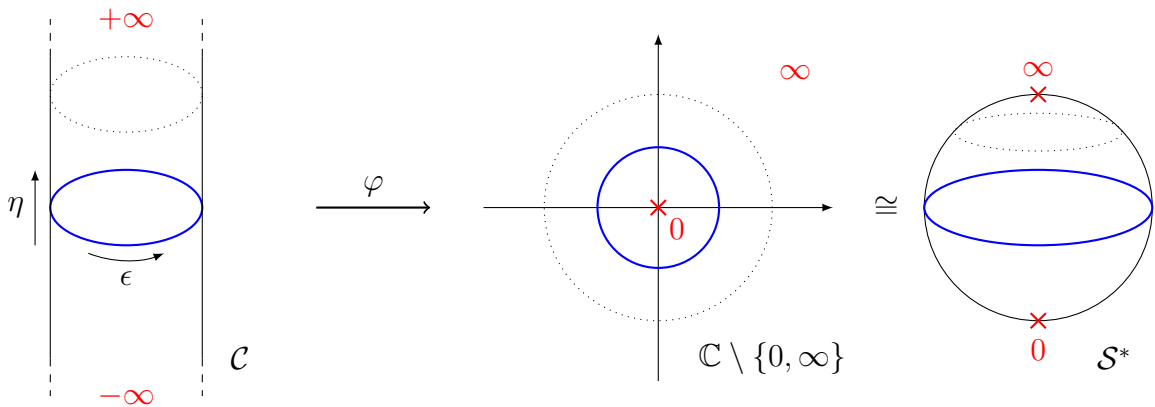

We would like to apply the general theory discussed above for $Z=\exp{(-\ii \ep)}$.
To do so we first need to extend the real variable $\ep$ to the complex plane.
The phases $\ep$ being defined modulo $2\pi$, a natural extension of their domain of definition is the complex cylinder
$ \mathcal C = \{ z = \epsilon + \ii \eta \, | \, (\epsilon,\eta) \in  S_1 \times \mathbb R \}$, as depicted in figure \ref{fig:RiemSphere}.
Then we define the map
\begin{equation}\label{defvarphi}
 \varphi : \left\lbrace \begin{array}{lll}
            \mathcal C & \longrightarrow & \mathbb C \\
	    z & \longmapsto & \ee^{-\ii z} 
           \end{array}\right.
\end{equation}
that sends this cylinder to the complex plane by preserving the circles (see Fig. \ref{fig:RiemSphere}). 
However, note that this map is not analytic since it has two essential singularities at $\eta \rightarrow +\infty$ and $\eta \rightarrow -\infty$,
mapped respectively to $\infty$ and $0$ in $\mathbb C$. 
This is specific to unitary models where the phase is $U(1)$-valued whereas such
singularities do not appear when doing $E \mapsto z$ for real energy $E$ of Hermitian systems \cite{HatsugaiPRB93}. 
These singularities will stay all along the construction and will be actually necessary.
Indeed, the image of $\mathcal C$ by $\varphi$ is  the Riemann sphere $\mathcal S^*$ \textit{punctured} of two singular points $0$ and $\infty$.  
This surface is not compact anymore, but outside this two points the function is still analytic such that around the image of $U(1)$
 (the equator in this picture), phase terms $\exp(-\ii \epsilon)$ can be extended in an analytic way.
 
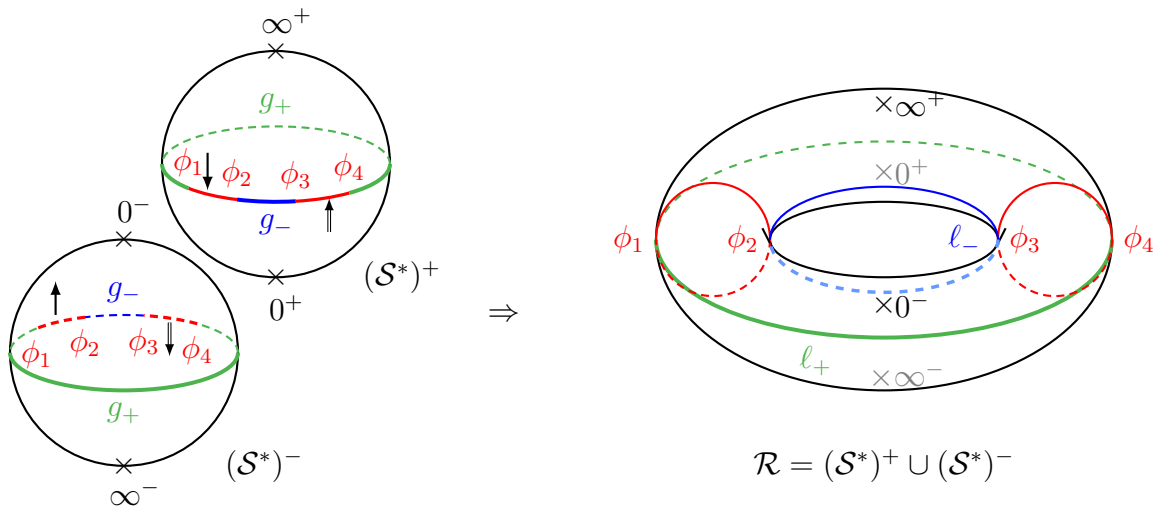
\begin{figure}[htb]
 \centering
\begin{tikzpicture}

\begin{scope}[yshift=2cm]
\draw[thick] (0,0) circle (1.5);
\draw[thick,densely dashed,vert] (0:1.5 and 0.5) arc (0:180:1.5 and 0.5);
\draw[vert,line width=1.5pt] (0:1.5 and 0.5) arc (0:-50:1.5 and 0.5);
\draw[blue,line width=1.5pt] (-80:1.5 and 0.5) arc (-80:-110:1.5 and 0.5);
\draw[vert,line width=1.5pt] (-140:1.5 and 0.5) arc (-140:-180:1.5 and 0.5);

\draw[red,very thick] (-50:1.5 and 0.5)node[above]{$\phi_4$} arc (-50:-80:1.5 and 0.5)node[above]{$\phi_3$};
\draw[red,very thick] (-110:1.5 and 0.5)node[above]{$\phi_2$} arc (-110:-140:1.5 and 0.5)node[above]{$\phi_1$};

\draw (1,-1.5) node[right]{$(\mathcal S^*)^+$};
\draw[blue] (0,-0.8)node{$g_-$};
\draw[vert] (0,0.8) node{$g_+$};
\draw (0,1.5) node {$\times$};
\draw (0.14,1.9) node{$\infty^+$};
\draw (0,-1.5) node {$\times$};
\draw (0.14,-1.9) node{$0^+$};

 \draw[-latex,thick] (-0.9,0.15) -- (-0.9,-0.35);
 \draw[-latex,double] (0.7,-0.9) -- (0.7,-0.45);
\end{scope}

\begin{scope}[yshift=-0.5cm,xshift=-2cm]
\draw[thick] (0,0) circle (1.5);
\draw[line width=1.5pt,vert] (0:1.5 and 0.5) arc (0:-180:1.5 and 0.5);
\draw[thick,densely dashed,vert] (0:1.5 and 0.5) arc (0:50:1.5 and 0.5);
\draw[thick,densely dashed,blue] (80:1.5 and 0.5) arc (80:110:1.5 and 0.5);
\draw[thick,densely dashed,vert] (140:1.5 and 0.5) arc (140:180:1.5 and 0.5);
\draw[red,densely dashed,very thick] (50:1.5 and 0.5)node[below]{$\phi_4$} arc (50:80:1.5 and 0.5)node[below]{$\phi_3$};
\draw[red,densely dashed,very thick] (110:1.5 and 0.5)node[below]{$\phi_2$} arc (110:140:1.5 and 0.5)node[below]{$\phi_1$};

\draw (2.5,-1.5) node[left]{$(\mathcal S^*)^-$};
\draw[blue] (0,0.8)node{$g_-$};
\draw[vert] (0,-0.8) node{$g_+$};

\draw (0,1.5) node {$\times$};
\draw (0.14,1.9) node{$0^-$};
\draw (0,-1.5) node {$\times$};
\draw (0.14,-1.55) node[below] {$\infty^-$};

\draw[-latex,thick] (-0.9,0.5) -- (-0.9,0.95) ;
\draw[-latex,double] (0.6,0.4) -- (0.6,-0.05);
\end{scope}

\begin{scope}[xshift=8cm,yshift=1cm]
 \draw[thick] (0,0) ellipse (3 and 2);
 \draw[vert,dashed,thick] (0:3 and 1.3) arc (0:180:3 and 1.3);
 \draw[vert,line width=1.5pt] (0:3 and 1.3) arc (0:-180:3 and 1.3);
 \draw[thick] (0,0) ellipse (1.5 and 0.5);
 \draw[red,thick] (-1.5,0)node[left]{$\phi_2$} arc (0:180:0.75)node[left]{$\phi_1$};
 \draw[red,thick,densely dashed] (-1.5,0) arc (0:-180:0.75);
 \draw[red,thick] (3,0)node[right]{$\phi_4$} arc (0:180:0.75)node[right]{$\phi_3$};
 \draw[red,thick,densely dashed] (3,0) arc (0:-180:0.75);
  
\draw[thick] (1.5,-0.05) -- (1.6,0.15); \draw[thick] (-1.5,-0.05) -- (-1.6,0.15);

\draw[blue,line width=0.8pt] (0:1.5 and 0.7) arc (0:180:1.5 and 0.7);
\draw[bleu,line width=1.3pt,dashed] (0:1.5 and 0.7) arc (0:-180:1.5 and 0.7);

\draw (0,-3) node {$\mathcal R = (\mathcal S^*)^+ \cup ( \mathcal S^*)^-$};
\draw (0,-0.87) node{$\times$};
\draw (0,-0.87) node[right]{$0^-$};
\draw[gray] (0,0.9) node{$\times$};
\draw[gray] (0,0.9) node[right]{$0^+$};
\draw[blue] (0.7,0) node[right]{$\ell_-$};
\draw[gray] (0.3,-1.80) node{$\times \infty^-$};
\draw[vert] (-0.9,-1.6) node{$\ell_+$};
\draw (0,1.80) node{$\times$};
\draw (0,1.80) node[right]{$\infty^+$};
\end{scope}

\draw (3,0) node{$\Rightarrow$};

\end{tikzpicture}
\caption{\small Construction of the Riemann surface $\mathcal R$ by gluing two punctured Riemann spheres
$(\mathcal S^*)^+$ and $(\mathcal S^*)^-$.
The branch cuts, in red, correspond to the bands. 
They and are delimited by the $\phi_i$ according to Eq. \eqref{eq:muQe} and separate  
the two gap $g_-$ (blue) from the gap $g_+$ (green). 
For convenience, we have flipped one sphere in a way such that when arriving at a branch cut we are going to
the other surface, stay to the same hemisphere (north or south), as illustrated by the arrows.
The gaps form two non-contractile loops $\ell_-$ and $\ell_+$ on $\mathcal R$ which are non-homotopic one to each other 
because of the essential singularities $\infty^{\pm}$ and $0^{\pm}$. 
\label{fig:RiemSurf}}
\end{figure}

One can now apply the general method of the previous section to construct the Riemann surface by replacing the two
cut Riemann spheres $\mathcal S^{\pm}$ by two punctured cut Riemann spheres $(\mathcal S^*)^{\pm}$.
The Riemann surface $\mathcal R$, defined as some pull-back of $\mathcal R_0$ by $\varphi$,
\begin{equation}
\mathcal R  = \left\lbrace (\mu, z) \in \mathbb C \times \mathcal C \, \left| \,
(\mu, Z = \varphi(z)=\ee^{-\ii z}) \in \mathcal R_0 \right.\right\rbrace 
\end{equation}
is obtained by gluing $(\mathcal S^*)^+$ and $(\mathcal S^*)^-$  together 
by the branch cuts $W(\ee^{-\ii z}) < 0$ which correspond to the regions of the two bands $\ep$ (at fixed $k$) as shown in Fig. \ref{fig:RiemSurf}.
These bands, together with the gaps $g_{\pm}$, constitute the equator of each punctured sphere as they originate from the 
real part of the complex variable $z$ (blue circle in figure \ref{fig:RiemSphere}).
All the quantities used, besides the square root, only involve polynomials in $\ee^{-\ii z}$ and $\ee^{\ii z}$ that are perfectly smooth on $\mathcal R$
 since  they are so on each copy of $\mathcal S^*$.
The square root of $\mu_k^2$ is an analytic function on the Riemann surface $\mathcal R$ and so are the eigenvalues of the transfer matrix. 
In particular, one can write $\lambda =f + \mu $ that corresponds to $\lambda_+$ when $\ee^{-\ii z}$ is in a gap
and the corresponding $\mu$ is positive, and similarly for $\lambda_-$ with $\mu$ negative. In this picture $\lambda$ contains
both square roots and one can go analytically from one to the other using the continuation in the complex numbers.
Consequently the edge states of the system are now described
in an analytic way on $\mathcal R$, and the full square root is given by one single formula instead of two.

Geometrically, the Riemann surface $\mathcal R$ is a torus with four punctured points (see Fig. \ref{fig:RiemSurf}).  
Each gap is present in two copies that generates loops along the torus.
Indeed, the two copies of $g_+$ generate the outer loop $\ell_+$ (in green) and the two copies of $g_-$ generate the inner loop $\ell_-$ (in blue). 
Importantly, due to the four singular points $0^+, \infty^+, 0^-$ and $\infty^-$ that cannot be crossed, 
these two loops are not equivalent (or homotopic) as in the standard torus since they cannot 
be deformed one to each other without crossing a singular point.
Hence the Riemann surface $\mathcal R$ keeps track of the two distinct gaps $g_-$ and $g_+$ of the unitary problem. 
Finally note that such relative position (interior/exterior) is completely arbitrary and just depends on the way we 
draw the construction of the torus. Indeed the two configurations of torus are topologically equivalent by "twisting" the full torus.

The  construction of the Riemann surface was performed at fixed quasi-momentum $k$.
When varying $k$, the size of the gaps changes so that one obtains a family of Riemann surfaces $\mathcal R_k$.
Still, the topology remains the same as these surfaces are all homotopic one to another (as long as the gaps do not close): one can deform all this family to the same torus $\mathcal R$ for all $k$.

\subsection{Winding number}

When $k$ spans $S_1$, an edge state may cross a gap by moving from one band to the other one, whereas the other edge state, 
located at the other boundary, crosses the gap in the opposite direction. 
On the Riemann torus $\mathcal R$, these two edge states span one of the two loops $\ell_{\pm}$ and can thus be qualified as \textit{topological}:
the winding of their pair cannot change value unless the gap closes.
This defines a topological invariant \textit{for each gap} that counts algebraically the number of chiral edge states.
As noticed by Hatsugai \cite{HatsugaiPRB93}, this winding number is nothing but the intersection number of the curve spanned by the edge state dispersion relation
$\ep(k)$ (after having identified all the Riemann surfaces $\mathcal{R}_k$) and some "vertical line" of the torus, e.g. one of the branch cuts.

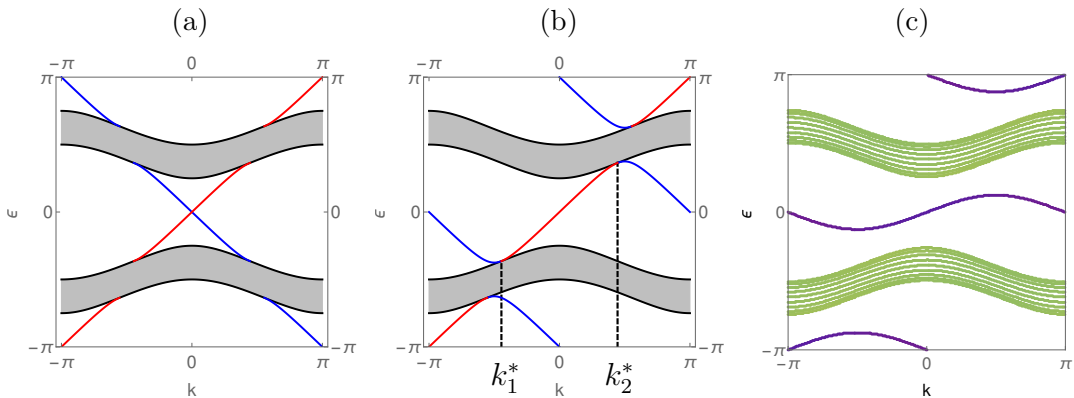
\begin{figure}[ht]
\begin{tikzpicture}
\centering
\node (0,0){ \includegraphics[scale=0.37]{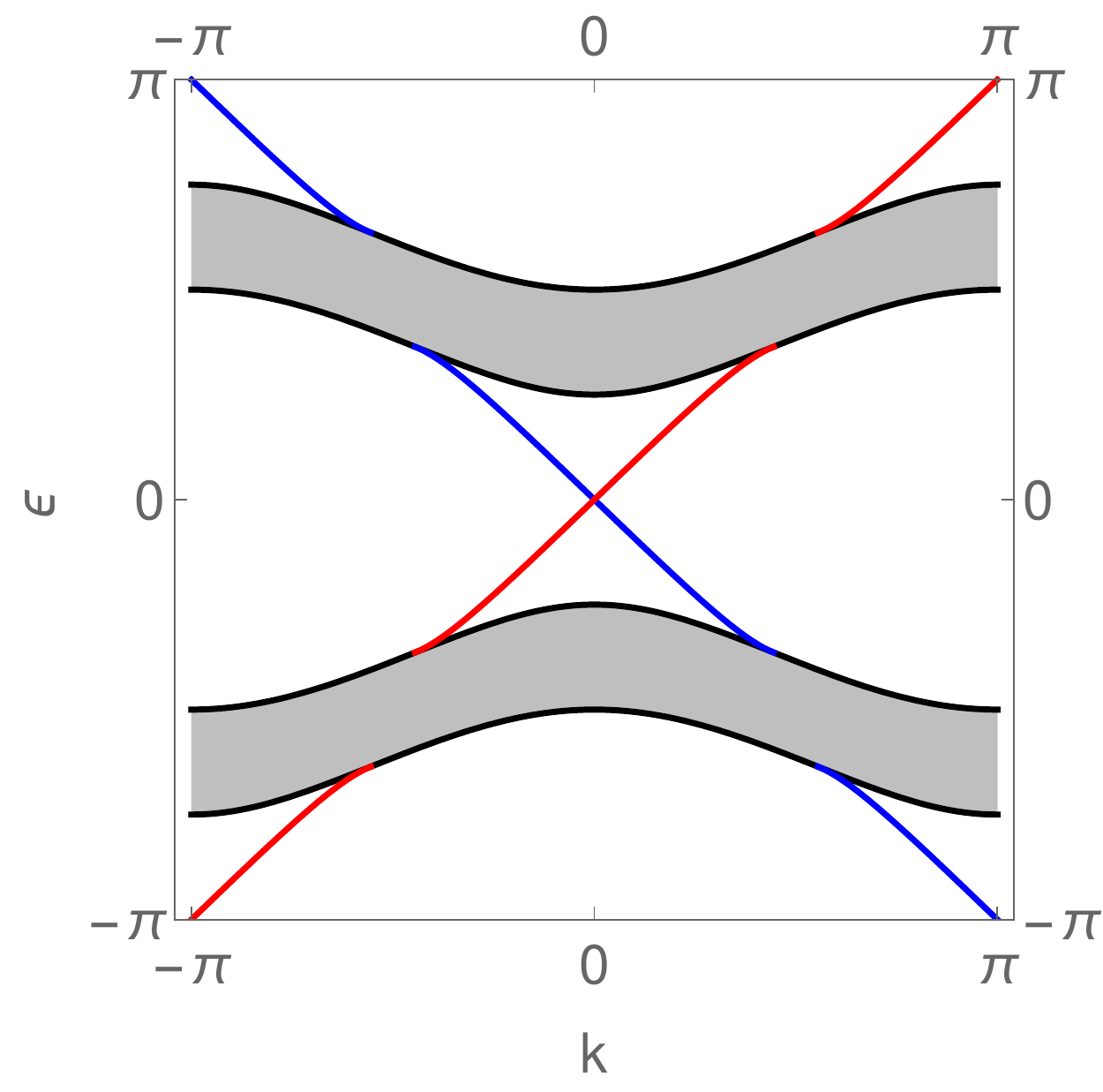}
 \includegraphics[scale=0.37]{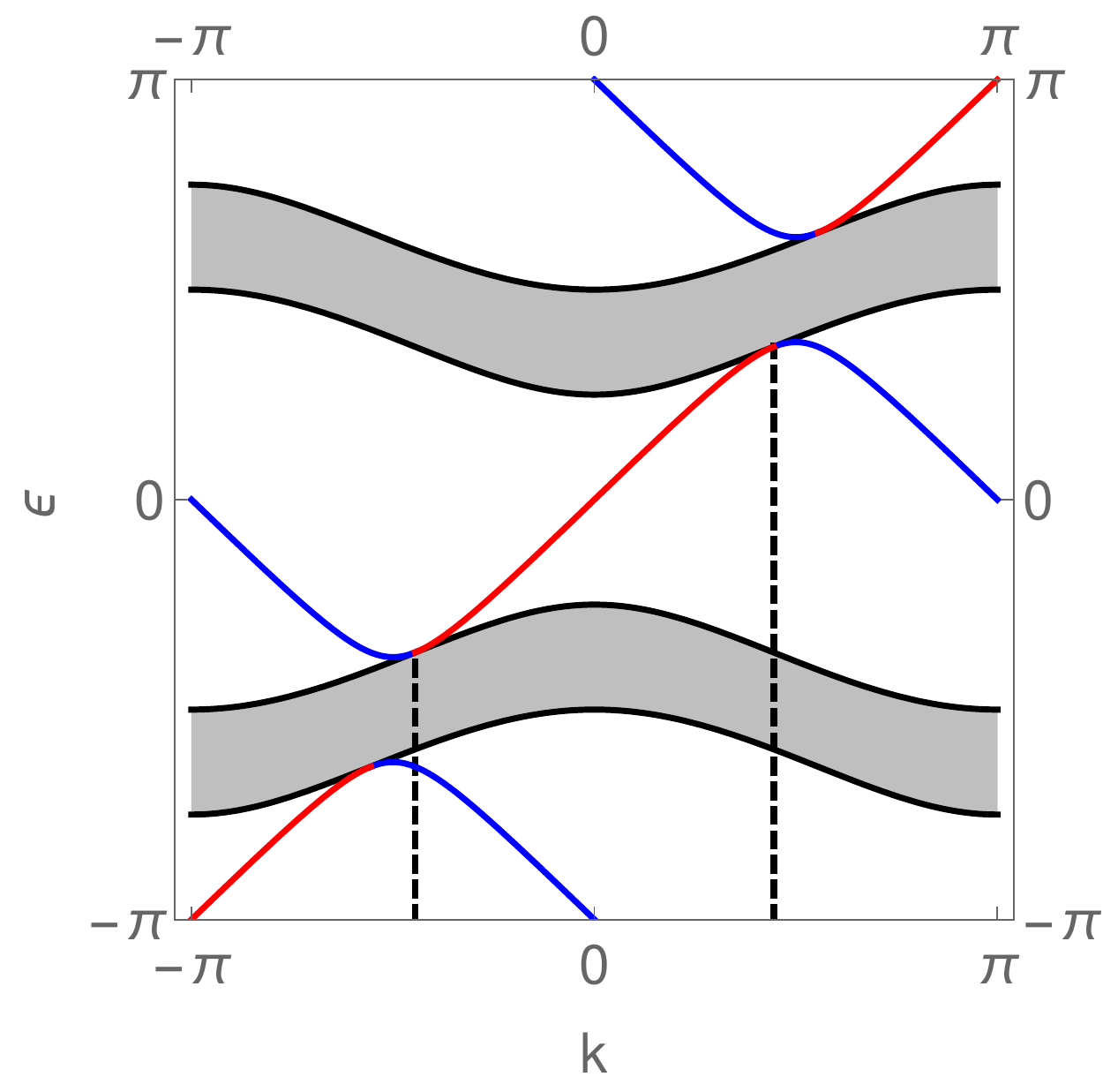}
  \includegraphics[scale=0.35]{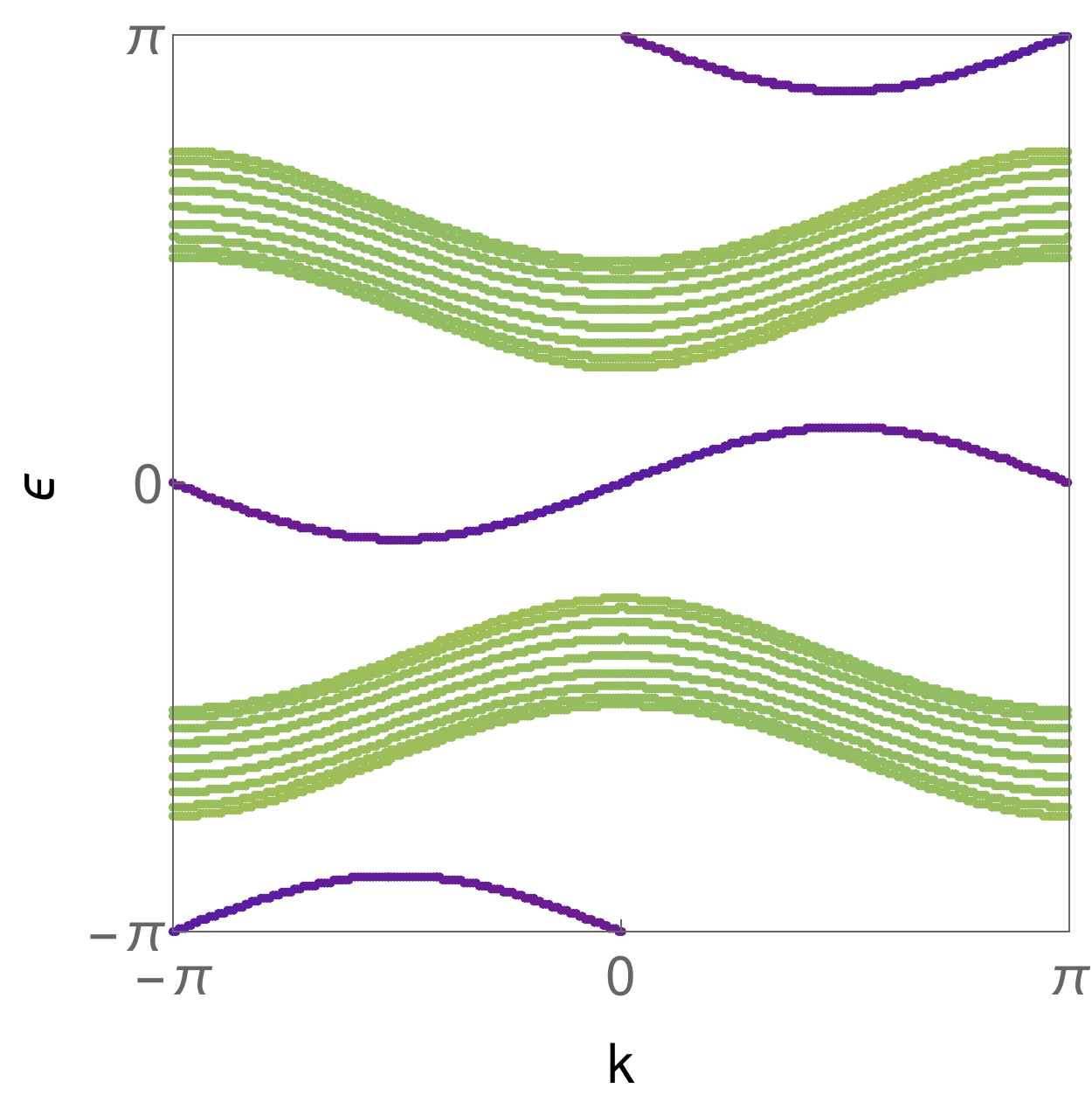}};
  \draw (-0.4,-2) node{$k_1^*$};
  \draw (1.1,-2) node{$k_2^*$};
  \draw (-4.55,2.7) node{\small (a)};
  \draw (0.3,2.7) node{\small (b)};
  \draw (4.95,2.7) node{\small (c)};
  \end{tikzpicture}
  \caption{\small{(a) Bulk bands and edge states obtained in the thermodynamic limit with the transfer matrix method for $\theta=\pi/8$. 
 (b) The boundary state at edge $n=1$ (blue) has been shifted ($\varphi=\pi$) so that the two edge states touch at $k_1^*$ and $k_2^*$.
 The dispersion relation of a pair of edge states (in each gap) is single-valued. (c) Appearance of trivial edge states for $\theta=3\pi/8$
 when complex phases are added to the model \eqref{eq:model}.}}
 \label{fig:shifted}
  \end{figure}

To compute explicitly this number $W$, it is particularly convenient to deal with one continuous function $\tilde{\ep}(k)$ for the dispersion relation of the pair of edge states 
rather than a multi-valued function as it is the case when the two edge states cross in the gap (see figure \ref{fig:shifted} (a)). 
Such a single-valued function can always be obtained by shifting one of the two edge state's dispersion relation, as shown in figure \ref{fig:shifted} (b).
This is performed by adding a phase-shift to the boundary vector (e.g. $\bone$) coefficients as $\tau'_1\rightarrow \tau'_1\ee^{\ii\varphi}$ and $\rho'_1\rightarrow \rho'_1\ee^{\ii\varphi}$
so that the unitarity of $U_1$ (and then $\tilde{U}$) is preserved\footnote{Similarly the unitary transformation $\tau_1\rightarrow \tau_1\ee^{\ii\varphi}$ and $\rho_1\rightarrow \rho_1\ee^{\ii\varphi}$
shifts the other edge state's dispersion relation.}. 
Importantly, the transfer matrix is not affected by this transformation, hence both the bulk bands and the construction of the Riemann surface remain the same.

It is enough to focus on the gap $g_-$ only.
The single-valued dispersion relation $\tilde{\ep}_-(k)$ of the pair of edge states is smooth and periodic, so that the winding number of $\exp(-\ii\tilde{\ep}_-(k))$ always vanishes.
However in the Riemann surface picture it can be used to wind around nontrivial loops. 
In this gap, $\tilde {\ep}_-(k)$ touches the bands in $k_1^*$ and $k_2^*$,
that is $\tilde{\ep}_-(k_1^*) = G_{-,t}(k_1^*)$ and $\tilde{\ep}_-(k_2^*) = G_{-,b}(k_2^*)$, for $G_{-,t/b}$ given in \eqref{eq:gap_edge}. 
That way, one gets $\mu_k >0$ for $k\in ]k_1^*,k_2^*[$ and $\mu_k <0$ for $k\in ]0,k_1^*[\, \cup \, ]k_2^*,\pi[$. 
Let us then define the function
\begin{equation}\label{defFD}
F_-(k)\equiv \rm{sign}(\mu_k) D_-(k)
\end{equation}
where $D_-(k)$ measures the relative distance between the dispersion relation of the edge state
 and a bound of the gap $g_-$. Let us choose the lower bound as a convention\footnote{Note that the other convention would reverse the sign of the winding number.};
in that case, $D_-(k)=\left(\tilde{\ep}_-(k)-G_{-,b}(k)\right)/\left(G_{-,t}-G_{-,b}(k)\right)$.
The function $F_-$ is well defined as soon as the gap does not close ($\theta \neq \pi/4$). 
It is $2\pi$-periodic in $k$, continuous in $k_1^*$ (since it is $0$) but not continuous at $k_2^*$. 
However the function
\begin{equation}
w_-(k) = \ee^{-\ii \pi F_-(k)}
\end{equation}
is continuous, and has non trivial winding number 
 \begin{equation}
W[w_-] = \dfrac{1}{2\pi \ii} \int_{-\pi}^{\pi} w_-^{-1}(k) \dd w_-(k) =  -\dfrac{1}{2} \int_{-\pi}^{\pi} \dd F_-(k)\ .
\end{equation}
Since $F_-$ is periodic, then, if $F_-$ were also continuous, its winding number would be $0$. 
However $F_-$ is only piecewise continuous, such that
\begin{align}
W[w] &= -\dfrac{1}{2} \int_{k_1^*}^{k_2^*} \dd F(k)- \dfrac{1}{2} \int_{k_2^*}^{k_1^*+2\pi} \dd F(k) 
 \, =  D(k_1^*) - D(k_2^*) 
 =  1 \label{eq:Ww} 
\end{align}
where we have dropped the gap index because of the gap correspondence discussed in section \ref{sec:gap_correspondence}.
From an effective point of view, the computation of $W[w]$ amounts to count the number of times $\partial_k \mu_k$ changes sign at points $k_i^*$
such that $F(k_i^*)$ is discontinuous, that is $D(k_i^*)=1$. On can thus write $W[w]$ as an intersection number
\begin{equation}
 W[w] = \sum_{k_i^* \text{s.t.} D(k_i^*) = 1} \text{sign} \left.\dfrac{\partial \mu_k}{\partial k} \right|_{k=k_i^*}
\end{equation}
where the derivative $\partial_k \mu_k$ is well defined on the Riemann surface. 
In practice, when going from a lower edge state ($\mu_k <0$) to an upper one ($\mu_k >0$) 
while increasing $k$, the derivative of $\mu_k$ is positive and the winding number increases by $1$,  
and conversely decreases by $1$ when going from the upper to the lower Riemann sheet. 

It is clear that this analysis is still valid for several edge states and without the gap correspondence;
the invariant $W$ thus counting the number of topological edge states in a given gap.
Besides, it also distinguishes topological edge states from non-topological "accidental" edge states that may appear
for certain set of parameters, as illustrated in figure \ref{fig:shifted} (c). 
Indeed these edge states are associated to contractile loops on $\mathcal{R}$ and thus do not benefit of any topological robustness.
Their winding $\eqref{eq:Ww}$ is clearly zero.

Finally note that the loops around the essential singularities of $\mathcal R$ might be related to previous invariant $W$. For example, a loop around $0^-$ on Figure \ref{fig:RiemSurf} is homeotopic to the one passing by $\phi_1 \rightarrow \phi_2 \rightarrow \phi_3 \rightarrow \phi_4 \rightarrow \phi_1$ and traveling along left solid red curve, dashed blue part of $\ell^-$, right solid red curve and solid green part of $\ell^+$, respectively. This loop corresponds to a global path in the spectrum, crossing both bands and both gaps but with opposite localization on the edges. In particular, the sum of loops around facing singularities from distinct Riemann sheet (e.g. $0^-$ and $\infty^+$) are homeotopic to $\ell^+ \cup \ell^-$ and hence related to gap invariant $W$.

\section{Application to physical models}
\label{sec:models}

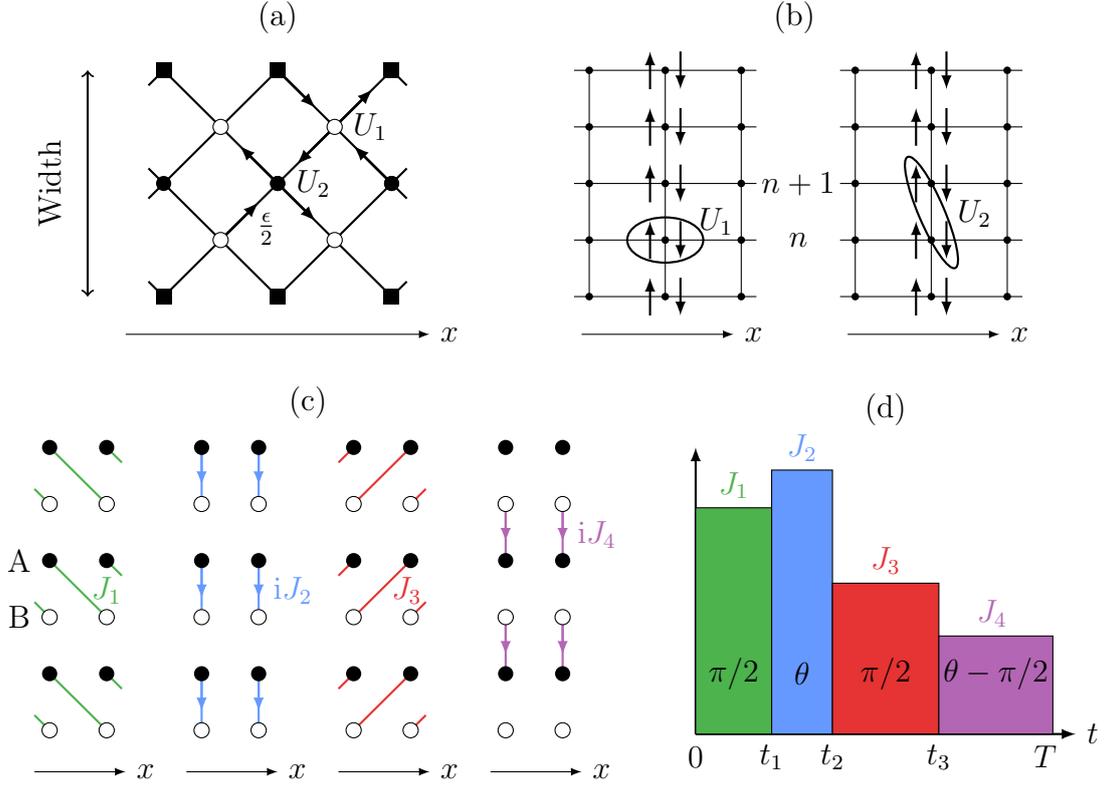
\begin{figure}[t]
\centering
\begin{tikzpicture}

\draw (3,3.7) node{(a)};

\draw[thick,<->] (0.5,0)--(0.5,3);

\draw (0,1.5) node[rotate=90]{Width};

\draw[thick] (1.3,1.3) -- (3,3);
\draw[thick] (1.5,0) -- (4.5,3);
\draw[thick] (3,0) -- (4.7,1.7);
\draw[thick] (4.5,0) -- (4.7,0.2);

\draw[thick] (1.3,1.7) -- (3,0);
\draw[thick] (1.5,3) -- (4.5,0);
\draw[thick] (3,3) -- (4.7,1.3);

\draw[thick] (1.3,0.2) -- (1.5,0);
\draw[thick] (1.3,2.8) -- (1.5,3);
\draw[thick] (4.5,3)   -- (4.7,2.8);
  
\fill (1.4,-0.1) rectangle (1.6,0.1);
\fill (2.9,-0.1) rectangle (3.1,0.1);
\fill (4.4,-0.1) rectangle (4.6,0.1);

\fill (1.5,1.5)  circle (0.1);
\fill (3,1.5)    circle (0.1);
\fill (4.5,1.5)  circle (0.1); \draw (3.1,1.5) node[right]{$U_2$} ;

\fill (1.4,2.9) rectangle (1.6,3.1);
\draw (2.85,0.9) node{$\frac{\epsilon}{2}$};
\fill (2.9,2.9) rectangle (3.1,3.1);

\fill (4.4,2.9) rectangle (4.6,3.1);

\draw[-latex,line width=0.99pt] (3.75,2.25) -- (4.25,2.75);
\draw[-latex,line width=0.99pt] (3.75,2.25) -- (3.25,1.75);

\draw[-latex,line width=0.99pt] (4.5,1.5) -- (4,2);
\draw[-latex,line width=0.99pt] (3,3) -- (3.5,2.5);

\draw[-latex,line width=0.99pt] (3,1.5) -- (2.5,2);
\draw[-latex,line width=0.99pt] (3,1.5) -- (3.5,1);
\draw[-latex,line width=0.99pt] (2.25,0.75) -- (2.75,1.25);

\fill[white] (2.25,0.75) circle (0.1); \draw (2.25,0.75) circle (0.1);
\fill[white] (3.75,0.75) circle (0.1); \draw (3.75,0.75) circle (0.1);

\fill[white] (2.25,2.25) circle (0.1); \draw (2.25,2.25) circle (0.1);
\fill[white] (3.75,2.25) circle (0.1); \draw (3.75,2.25) circle (0.1);

\draw (3.85,2.25) node[right]{$U_1$} ;

\fill (3,1.5) circle (0.1);
\fill (4.5,1.5) circle (0.1);

\draw[-latex] (1,-0.5)--(5,-0.5) node[right]{$x$};

\begin{scope}[xshift=10cm]

\draw (-0.2,3.7) node{(b)};

 \draw (-1.9,0) -- (-1.9,3); 
 \fill (-1.9,0) circle (0.05);
 \fill (-1.9,0.75) circle (0.05);
 \fill (-1.9,1.5) circle (0.05);
 \fill (-1.9,2.25) circle (0.05);
 \fill (-1.9,3) circle (0.05);
 
 \draw (-0.9,0) -- (-0.9,3); 
 \fill (-0.9,0) circle (0.05);
 \fill (-0.9,0.75) circle (0.05);
 \fill (-0.9,1.5) circle (0.05);
 \fill (-0.9,2.25) circle (0.05);
 \fill (-0.9,3) circle (0.05);

 \draw (-2.9,0) -- (-2.9,3); 
 \fill (-2.9,0) circle (0.05);
 \fill (-2.9,0.75) circle (0.05);
 \fill (-2.9,1.5) circle (0.05);
 \fill (-2.9,2.25) circle (0.05);
 \fill (-2.9,3) circle (0.05);

 \draw (-3.1,3)--(-0.7,3);
 \draw (-3.1,2.25)--(-0.7,2.25);
 \draw (-3.1,1.5)--(-0.7,1.5);
\draw (-3.1,0.75)--(-0.7,0.75);
\draw (-3.1,0)--(-0.7,0);

\draw (-0.15,0.75) node{$n$};
\draw (-0.15,1.5) node{$n+1$};

 \draw[-latex,thick] (-2.1,-0.25)--(-2.1,0.25);
 \draw[-latex,thick] (-1.7,0.25)--(-1.7,-0.25);

 \draw[-latex,thick] (-2.1,0.5) -- (-2.1,1);
 \draw[-latex,thick] (-1.7,1) -- (-1.7,0.5);
 
 \draw[-latex,thick] (-2.1,1.25) -- (-2.1,1.75);
 \draw[-latex,thick] (-1.7,1.75) -- (-1.7,1.25);
 
 \draw[-latex,thick] (-2.1,2) -- (-2.1,2.5);
 \draw[-latex,thick] (-1.7,2.5) -- (-1.7,2);
 
 \draw[-latex,thick] (-2.1,2.75) -- (-2.1,3.25);
 \draw[-latex,thick] (-1.7,3.25) -- (-1.7,2.75);

 \draw[thick] (-1.9,0.75) ellipse (0.5 and 0.3);

 \draw (-1.6,1) node[right]{$U_1$};
\draw[-latex] (-3,-0.5)--(-1,-0.5) node[right]{$x$}; 
 \end{scope}
 
 \begin{scope}[xshift=13.5cm]
 
 \draw (-1.9,0) -- (-1.9,3); 
 \fill (-1.9,0) circle (0.05);
 \fill (-1.9,0.75) circle (0.05);
 \fill (-1.9,1.5) circle (0.05);
 \fill (-1.9,2.25) circle (0.05);
 \fill (-1.9,3) circle (0.05);
 
 \draw (-0.9,0) -- (-0.9,3); 
 \fill (-0.9,0) circle (0.05);
 \fill (-0.9,0.75) circle (0.05);
 \fill (-0.9,1.5) circle (0.05);
 \fill (-0.9,2.25) circle (0.05);
 \fill (-0.9,3) circle (0.05);

 \draw (-2.9,0) -- (-2.9,3); 
 \fill (-2.9,0) circle (0.05);
 \fill (-2.9,0.75) circle (0.05);
 \fill (-2.9,1.5) circle (0.05);
 \fill (-2.9,2.25) circle (0.05);
 \fill (-2.9,3) circle (0.05);

 \draw (-3.1,3)--(-0.7,3);
 \draw (-3.1,2.25)--(-0.7,2.25);
 \draw (-3.1,1.5)--(-0.7,1.5);
\draw (-3.1,0.75)--(-0.7,0.75);
\draw (-3.1,0)--(-0.7,0);

 \draw[-latex,thick] (-2.1,-0.25)--(-2.1,0.25);
 \draw[-latex,thick] (-1.7,0.25)--(-1.7,-0.25);

 \draw[-latex,thick] (-2.1,0.5) -- (-2.1,1);
 \draw[-latex,thick] (-1.7,1) -- (-1.7,0.5);
 
 \draw[-latex,thick] (-2.1,1.25) -- (-2.1,1.75);
 \draw[-latex,thick] (-1.7,1.75) -- (-1.7,1.25);
 
 \draw[-latex,thick] (-2.1,2) -- (-2.1,2.5);
 \draw[-latex,thick] (-1.7,2.5) -- (-1.7,2);
 
 \draw[-latex,thick] (-2.1,2.75) -- (-2.1,3.25);
 \draw[-latex,thick] (-1.7,3.25) -- (-1.7,2.75);

 \draw[shift={(-1.5,-0.78)},thick,rotate=-67] (-1.9,0.375) ellipse (0.8 and 0.17);
  
 \draw (-1.7,1.1) node[right]{$U_2$};
\draw[-latex] (-3,-0.5)--(-1,-0.5) node[right]{$x$}; 
  
\end{scope}

\begin{scope}[yshift=-2cm]

\draw[vert] (0.75,-1.9) node{$J_1$};

\draw (-0.4,-1.5) node{A};
\draw (-0.4,-2.25) node{B};
\draw[thick,vert] (-0.2,-0.55) -- (0,-0.75); \draw[thick,vert] (0.75,0)--(0.95,-0.2);
\draw[thick,vert] (-0.2,-2.05) -- (0,-2.25); \draw[thick,vert] (0.75,-1.5)--(0.95,-1.7);
\draw[thick,vert] (-0.2,-3.55) -- (0,-3.75); \draw[thick,vert] (0.75,-3)--(0.95,-3.2);
\draw[thick,vert] (0,0) -- (0.75,-0.75);
\draw[thick,vert] (0,-1.5) -- (0.75,-2.25);
\draw[thick,vert] (0,-3) -- (0.75,-3.75);

\fill (0,0) circle (0.1);    
\fill (0.75,0) circle (0.1);

\fill (0,-1.5) circle (0.1);
\fill (0.75,-1.5) circle (0.1);

\fill (0,-3) circle (0.1);
\fill (0.75,-3) circle (0.1);

\fill[white] (0,-0.75) circle (0.1); \draw (0,-0.75) circle (0.1);
\fill[white] (0.75,-0.75) circle (0.1); \draw (0.75,-0.75) circle (0.1);

\fill[white] (0,-2.25) circle (0.1); \draw (0,-2.25) circle (0.1);
\fill[white] (0.75,-2.25) circle (0.1); \draw (0.75,-2.25) circle (0.1);

\fill[white] (0,-3.75) circle (0.1); \draw (0,-3.75) circle (0.1);
\fill[white] (0.75,-3.75) circle (0.1); \draw (0.75,-3.75) circle (0.1);

\draw[-latex] (-0.2,-4.3)--(1,-4.3) node[right]{$x$}; 
\end{scope}

\begin{scope}[yshift=-2cm,xshift=2cm]

\draw (1.4,0.6) node{(c)};
\draw[bleu] (1.2,-1.9) node{$\ii J_2$};

\draw[thick,bleu] (0,0) -- (0,-0.75);
\draw[thick,-latex,bleu] (0,0) -- (0,-0.50);
\draw[thick,bleu] (0.75,0) -- (0.75,-0.75);
\draw[thick,-latex,bleu] (0.75,0) -- (0.75,-0.5);
\draw[thick,bleu] (0,-1.5) -- (0,-2.25);
\draw[thick,-latex,bleu] (0,-1.5) -- (0,-2);
\draw[thick,bleu] (0.75,-1.5) -- (0.75,-2.25);
\draw[thick,-latex,bleu] (0.75,-1.5) -- (0.75,-2);
\draw[thick,bleu] (0,-3) -- (0,-3.75);
\draw[thick,-latex,bleu] (0,-3) -- (0,-3.5);
\draw[thick,bleu] (0.75,-3) -- (0.75,-3.75);
\draw[thick,-latex,bleu] (0.75,-3) -- (0.75,-3.5);

\fill (0,0) circle (0.1);    
\fill (0.75,0) circle (0.1);

\fill (0,-1.5) circle (0.1);
\fill (0.75,-1.5) circle (0.1);

\fill (0,-3) circle (0.1);
\fill (0.75,-3) circle (0.1);

\fill[white] (0,-0.75) circle (0.1); \draw (0,-0.75) circle (0.1);
\fill[white] (0.75,-0.75) circle (0.1); \draw (0.75,-0.75) circle (0.1);

\fill[white] (0,-2.25) circle (0.1); \draw (0,-2.25) circle (0.1);
\fill[white] (0.75,-2.25) circle (0.1); \draw (0.75,-2.25) circle (0.1);

\fill[white] (0,-3.75) circle (0.1); \draw (0,-3.75) circle (0.1);
\fill[white] (0.75,-3.75) circle (0.1); \draw (0.75,-3.75) circle (0.1);

\draw[-latex] (-0.2,-4.3)--(1,-4.3) node[right]{$x$}; 
\end{scope}

\begin{scope}[yshift=-2cm,xshift=4cm]

\draw[rouge] (0.7,-1.9) node{$J_3$};

\draw[thick,rouge] (0.75,-0.75) -- (0.95,-0.55); \draw[thick,rouge] (-0.2,-0.2)--(0,0);
\draw[thick,rouge] (0.75,-2.25) -- (0.95,-2.05); \draw[thick,rouge] (-0.2,-1.7)--(0,-1.5);
\draw[thick,rouge] (0.75,-3.75) -- (0.95,-3.55); \draw[thick,rouge] (-0.2,-3.2)--(0,-3);

\draw[thick,rouge] (0.75,0) -- (0,-0.75);
\draw[thick,rouge] (0.75,-1.5) -- (0,-2.25);
\draw[thick,rouge] (0.75,-3) -- (0,-3.75);

\fill (0,0) circle (0.1);    
\fill (0.75,0) circle (0.1);

\fill (0,-1.5) circle (0.1);
\fill (0.75,-1.5) circle (0.1);

\fill (0,-3) circle (0.1);
\fill (0.75,-3) circle (0.1);

\fill[white] (0,-0.75) circle (0.1); \draw (0,-0.75) circle (0.1);
\fill[white] (0.75,-0.75) circle (0.1); \draw (0.75,-0.75) circle (0.1);

\fill[white] (0,-2.25) circle (0.1); \draw (0,-2.25) circle (0.1);
\fill[white] (0.75,-2.25) circle (0.1); \draw (0.75,-2.25) circle (0.1);

\fill[white] (0,-3.75) circle (0.1); \draw (0,-3.75) circle (0.1);
\fill[white] (0.75,-3.75) circle (0.1); \draw (0.75,-3.75) circle (0.1);

\draw[-latex] (-0.2,-4.3)--(1,-4.3) node[right]{$x$}; 
\end{scope}

\begin{scope}[yshift=-2cm,xshift=6cm]

\draw[violet2] (1.2,-1.15) node{$\ii J_4$};

\draw[thick,violet2] (0,-0.75) -- (0,-1.5);
\draw[thick,-latex,violet2] (0,-0.75) -- (0,-1.25);
\draw[thick,violet2] (0.75,-0.75) -- (0.75,-1.5);
\draw[thick,-latex,violet2] (0.75,-0.75) -- (0.75,-1.25);
\draw[thick,violet2] (0,-2.25) -- (0,-3);
\draw[thick,-latex,violet2] (0,-2.25) -- (0,-2.75);
\draw[thick,violet2] (0.75,-2.25) -- (0.75,-3);
\draw[thick,-latex,violet2] (0.75,-2.25) -- (0.75,-2.75);

\fill (0,0) circle (0.1);    
\fill (0.75,0) circle (0.1);

\fill (0,-1.5) circle (0.1);
\fill (0.75,-1.5) circle (0.1);

\fill (0,-3) circle (0.1);
\fill (0.75,-3) circle (0.1);

\fill[white] (0,-0.75) circle (0.1); \draw (0,-0.75) circle (0.1);
\fill[white] (0.75,-0.75) circle (0.1); \draw (0.75,-0.75) circle (0.1);

\fill[white] (0,-2.25) circle (0.1); \draw (0,-2.25) circle (0.1);
\fill[white] (0.75,-2.25) circle (0.1); \draw (0.75,-2.25) circle (0.1);

\fill[white] (0,-3.75) circle (0.1); \draw (0,-3.75) circle (0.1);
\fill[white] (0.75,-3.75) circle (0.1); \draw (0.75,-3.75) circle (0.1);

\draw[-latex] (-0.2,-4.3)--(1,-4.3) node[right]{$x$}; 
\end{scope}

\begin{scope} [yshift=-6cm,xshift=8.5cm]
 
 \draw (2.5,4.5) node{(d)};
 \draw[thick,-latex] (0,0.2)--(5,0.2) node[right]{$t$}; 
 \draw[thick,-latex] (0,0.2)--(0,4);
 \draw[fill=vert] (0,0.2) rectangle (1,3.2);
 \draw[fill=bleu] (1,0.2) rectangle (1.8,3.7);
 \draw[fill=rouge] (1.8,0.2) rectangle (3.2,2.2);
 \draw[fill=violet2] (3.2,0.2) rectangle (4.7,1.5);
 
 \draw (0,-0.1) node{$0$};
 \draw (1,-0.1) node{$t_1$};
 \draw (1.8,-0.1) node{$t_2$};
 \draw (3.2,-0.1) node{$t_3$};
 \draw (4.6,-0.1) node{$T$};
 
 \draw[vert] (0.5,3.5) node{$J_1$};
 \draw[bleu] (1.4,4) node{$J_2$};
 \draw[rouge] (2.5,2.5) node{$J_3$};
 \draw[violet2] (3.9,1.8) node{$J_4$};
 
 \draw (0.5,1) node{$\pi/2$};
 \draw (1.4,1) node{$\theta$};
 \draw (2.5,1) node{$\pi/2$};
 \draw (3.95,1) node{$\theta-\pi/2$};
 
\end{scope}

\end{tikzpicture}
\caption{\small (a-b-c) Three different physical models on a strip periodic in the $x$ direction described by the same
unitary matrix $\tilde{U}(k)$. 
(a) Oriented network where the white (black) nodes are characterized by scattering matrices $U_1$ ($U_2$).
The arrows represent the incoming and out-going states and a phase $\epsilon/2$ is accumulated between two adjacent nodes. 
The black squares constitute the boundaries where a total reflection occurs.  
(b) Spin-$1/2$ particles on a lattice on which a protocol of six steps is applied. 
Five of the steps are encoded into $U_1$ which is local in the transverse direction, whereas the rotation
$U_2$ is non-local. (c) Four steps of a periodic hopping process. 
The arrows represent a $\pi/2$ phase associated to the hopping. 
(d) Sketch of the pulses associated to each step of the cycle of period $T$. 
}
\label{fig:systems}
\end{figure}

As mentioned in section \ref{sec:two-gap}, the unitary matrix $\tilde{U}(k)$ defined in Eq. \eqref{eq:Ufactorized} maps on several physical two-dimensional
systems in a cylinder geometry. 
First, let us notice that from the factorized form \eqref{eq:Ufactorized}, it turns out that $\tilde{U}$ actually describes 
an oriented square lattice similar to the Ho-Chalker model \cite{Ho_Chalker96} as depicted in Fig.~\ref{fig:systems}.
In that case, $U_1$ and $U_2$ can be interpreted as scattering matrices that describe the coherent reflection and transmission processes at the nodes
of the network. It was shown by Chong and collaborators that such an oriented network actually also describes the propagation 
of electromagnetic modes in arrays of optical \cite{LiangPRL13, PasekChong14}
or micro-wave \cite{HuPRX15} resonators beyond the tight-binding model\footnote{Up to another choice than \eqref{eq:model} for the scattering parameters. 
 Despite a few technical differences the same method can be applied, leading to the same conclusions.}.

 
Interestingly, this model of a (static) network, also maps on other dynamical Floquet systems as we now show.
To see it explicitly, let us factorize $\tilde{U}(k)$ and replace the scattering parameters by their value (Eq.~\eqref{eq:model}).
One gets
\begin{equation}
 U_1(k) = \ee^{\ii \frac{\pi}{2} \sigma_x}\, \ee^{-\ii \frac{k}{2} \sigma_z}\, \ee^{\ii \theta \sigma_y}\,
          \ee^{-\ii \frac{\pi}{2} \sigma_x}\, \ee^{\ii \frac{k}{2} \sigma_z} \quad , \quad
U_2 = \ee^{-\ii (\theta-\pi/2)\sigma_y} \ .
    \label{eq:QWU1}
\end{equation}
Clearly, $\tilde{U}(k)$ reveals a quantum protocol that consists in six steps acting on a two-level system.
By repeating periodically this protocol, $\tilde{U}(k)$ can be interpreted as the Floquet operator (evolution operator after one period of time)
 of a discrete-time quantum walk (see Ref.~\cite{Kitagawa_QW} for a pedagogical introduction) of a quantum system consisting of spin-$1/2$ 
particles located at the nodes of a lattice. 
The first five steps are given by $\tilde{U}_1(k)$ which is
block-diagonal in the basis of the position across the cylinder's width. 
Thus, these operations are local in position as sketched in Fig.~\ref{fig:systems} (b) and correspond successively to various  shifts and spin-rotations.


Finally, the non block-diagonal operator $\tilde{U}_2$ is applied, so that the corresponding operation
(a spin-rotation by an angle $2\theta - \pi$ around the $y$ axis), is applied on a two-level
 quantum state which is delocalized on sites $n$ for spin down and $n+1$ for spin up.
During this last step, one spin at each edge of the strip is left unchanged.
In contrast with the oriented network model which is static, this describes a (Floquet) dynamical process.

This model can finally be mapped onto a time-dependent tight-binding model on a square lattice,
where the hopping amplitudes are successfully switched-on and off, in the spirit of previous Floquet toy models \cite{KitagawaPRB10,Rudner13,Carpentier15}
as illustrated in Fig.~\ref{fig:systems} (c).
 Unlike the two previous models, one can even write down explicitly the corresponding Bloch Hamiltonian 
 \begin{equation}
H(t,k) = \left\lbrace
  \begin{array}{llcl}
    J_1\, \mathbf{g}_+(k)\cdot \bm{\sigma}  &\equiv H_1(k) & \quad \text{for} \quad &   0 < t < t_1\\
    J_2 \sigma_y &\equiv H_2  & \quad\text{for}\quad &    t_1 < t < t_2 \\
    J_3\, \mathbf{g}_-(k) \cdot \bm{\sigma}&\equiv H_3(k)  &\quad \text{for}\quad &     t_2 < t < t_3 \\
    J_4 \sigma_y &\equiv H_4   & \quad\text{for}\quad &    t_3 <t< T
  \end{array}\right.
\label{eq:hamiltonian}
 \end{equation}
 where $\mathbf{g}_\pm(k) = \left(\cos{k/2},\pm \sin{k/2},0\right)$.
By giving a step profile to the time evolution of the couplings  as depicted in Fig.~\ref{fig:systems} (d),
then $U_1(k)$ and $U_2$ can be seen as the evolution operators 
\begin{equation}
 U_1(k) = \ee^{-\ii \int_{t_2}^{t_3}dt\, H_3(k)}\,\ee^{-\ii \int_{t_1}^{t_2}dt\, H_2}\,\ee^{-\ii \int_0^{t_1}dt\, H_1(k)} \quad , \quad U_2 = \ee^{-\ii \int_{T-t_3}^T dt\, H_4} \ .
\end{equation}
The Hamiltonian \eqref{eq:hamiltonian} has a simple interpretation sketched in Fig.~\ref{fig:systems} (c) and (d).
First, a coupling of amplitude $J_1$ is switched-on between second nearest neighbors (A and B sites).
Next, a purely imaginary intra-dimer coupling is switched-on with an amplitude $J_2$.
Then another coupling of amplitude $J_3$ is switched-on between second nearest neighbors. 
And finally, a purely imaginary inter-dimer coupling is switched-on with an amplitude $J_4$.
During this last step, sites at each side of the strip are left uncoupled\footnote{Notice that by imposing instead periodic 
boundary conditions, one gets the time ordered evolution operator for the bulk system from which a bulk gap topological invariant 
can be computed \cite{Rudner13,Carpentier15} and is found to be $-1$ 
($0$) in the topological (trivial) phase in agreement with our approach.}.
Considering that these four steps are repeated periodically in time, then the unitary matrix $\tilde{U}(k)$ 
is nothing but the Floquet operator in a cylinder geometry 
and the phase $\epsilon$ can thus be interpreted as the quasi-energy of the periodic dynamics, as for the previous model.  
 
These three models, described by the same unitary matrix $\tilde{U}(k)$ and thus the same transfer matrix, share the same topological properties
independently of whether a Hamiltonian or a periodic dynamics can be associated to them.
This illustrates the generality of the framework we have used along the paper.

\section{Discussion}

The transfer matrix approach allows the definition of a \textit{gap} topological invariant,
as opposed to a \textit{bulk} topological invariant defined for the bands of the periodic system, e.g. the Chern number.
This is particularly useful for unitary systems for which the Chern numbers can vanish while 
the system still exhibits topological edge states, as illustrated all along this paper.

The topological nature of the edge states is revealed by the Riemann surface when taking into account the boundary conditions.
Note that the construction of the Riemann surface only requires the transfer matrix that describes the (projected) bands in the thermodynamic limit.
It thus only contains information about the bulk (one-dimensional) system.
This interplay between the bulk information and the boundary conditions is encoded into 
the quantity $Q(k,\ep)$ (defined in Eq. \eqref{eq:domain_edgestate}) whose change of sign for every $\ep(k_0)$ in a gap at fixed $k_0$
guaranties the existence of an edge state, as shown in figure \ref{fig:edgestates}.  
It follows that a topological transition occurs when the gap closes at $(k_0,\ep_0)$
if and only if $Q(k_0,\ep_0)$ changes sign. 

Besides, the topological invariant $W$ appears explicitly as an obstruction in defining a continuous
and periodic argument for the phase eigenvalues $\ep$ on the underlying Riemann surface $\mathcal R$.
This invariant takes a geometrical meaning by counting the number of times a pair of edge states 
(located at two opposite edges) winds around one of the two loops $\ell_\pm$ of the punctured torus $\mathcal R$. 
The fact that these two loops are non-homotopic results from the existence of essential singularities that are absent in
Hermitian systems for which a similar construction was first performed \cite{HatsugaiPRB93}.

Other gap invariants have been proposed in two dimensions, especially in the context of Floquet systems which are described by a time ordered evolution operator \cite{Rudner13, Carpentier15}.
In that case, the additional time parameter allows for the definition of a gap topological invariant for the bulk (two dimensional) system, irrespective of
the boundary conditions.
In contrast, the approach developed in this paper applies for any unitary system in a finite size (cylindrical) geometry, irrespective of the existence
of a time dependent dynamics. Furthermore, other topological indexes associated to edge states of Hermitian systems have been already proposed
\cite{Schuba13, GrafPorta13, AgazziEckmannGraf14}.
Their generalization to unitary systems and the relation with the present index is a natural direction for future investigations.
Besides, it would be interesting to bridge the transfer matrix approach with the scattering matrix approach that also provides a 
topological characterization of edge states in both Hermitian \cite{Meidan11, FulgaPRB11} and  unitary \cite{Fulga15} systems. 
Unlike the transfer matrix which somehow probes a bulk property of the reduced (one-dimensional) system, the scattering matrix
requires to connect the system to a lead and thus only probes the edges properties.
Moreover, the winding number that is defined is assigned to one edge whereas the two edges are required in the Riemann torus picture.

Finally it would be very interesting to adapt this method to other exotic topological phases,
in particular when time-reversal symmetry \cite{Carpentier15, CarpentierNPB15} or dissipation \cite{Bardyn13, Budich15} plays a crucial role.

\textit{Acknowledgments}: This work was supported by the French Agence Nationale
de la Recherche (ANR) under grant TopoDyn (ANR-14-ACHN-0031). The authors thank M. Fruchart for his suggestions and fruitful discussions.

\bibliographystyle{unsrt}
\bibliography{biblio}

\end{document}